\documentclass[notitlepage,twocolumn,prx,nobalancelastpage,amssymb,superscriptaddress,showpacs]{revtex4-1}
\newsavebox{\mstrut}
\usepackage{amsmath}
\usepackage{soul}
\usepackage{hyperref}
\usepackage{float}
\usepackage{enumerate} 
\usepackage{mathtools}
\usepackage[usenames,dvipsnames]{xcolor}
\usepackage{graphicx}
\newsavebox\mysavebox
\usepackage{dcolumn}
\usepackage{bm}
\hypersetup{
    colorlinks=true,
    linkcolor=blue,
    filecolor=blue,      
    urlcolor=blue,
    citecolor=blue,
    linkcolor=red,
}
\usepackage{amsmath, amsthm, amssymb}
\usepackage{romannum}
\newcommand\norm[1]{\left\lVert#1\right\rVert}

\newcommand{\sidecaption}[1]

\newcommand{\bbra}[1]{%
    \sbox{\mstrut}{$#1$}%
    \mathinner{\left\langle\kern-0.5\ht\mstrut\left\langle{#1}\kern0.2\ht\mstrut\right|\right.}%
}
\newcommand{\kett}[1]{%
    \sbox{\mstrut}{$#1$}%
    \mathinner{\left|\left.{#1}\kern0.2\ht\mstrut\right\rangle\kern-0.5\ht\mstrut\right\rangle}%
}
\newcommand{\kettt}[1]{%
    \sbox{\mstrut}{$#1$}%
    \mathinner{\left|\left.{#1}\kern0.2\ht\mstrut\right\rangle\kern-0.7\ht\mstrut\right\rangle}%
}

\newcommand{\bbbra}[1]{%
    \sbox{\mstrut}{$#1$}%
    \mathinner{\left\langle\kern-0.7\ht\mstrut\left\langle{#1}\right|\right.}%
}

\newcommand{\ketttt}[1]{%
    \sbox{\mstrut}{$#1$}%
    \mathinner{\left|\left.{#1} \ \vphantom{\copy\mstrut}\right\rangle\kern-\dimexpr0.8\ht\mstrut\relax\right\rangle}%
}

\newcommand{\bbbbra}[1]{%
    \sbox{\mstrut}{$#1$}%
    \mathinner{\left\langle\kern-0.8\ht\mstrut\left\langle \ {#1}\kern0.2\ht\mstrut\right|\right.}%
}

\newcommand{\normsop}[1]{%
    \sbox{\mstrut}{$#1$}%
    \mathinner{\left\langle\kern-0.7\ht\mstrut\left\langle \ {#1}\kern0.2\ht\mstrut | {#1}\kern0.2\ht\mstrut\right\rangle\kern-0.7\ht\mstrut\right\rangle}%
}
\newcommand{\overlapsop}[2]{%
    \sbox{\mstrut}{$#1$}%
    \mathinner{\left\langle\kern-0.6\ht\mstrut\left\langle {#1}\kern0.2\ht\mstrut \big| {#2}\kern0.2\ht\mstrut\right\rangle\kern-0.6\ht\mstrut\right\rangle}%
}

\newcommand{\overlapso}[2]{%
    \sbox{\mstrut}{$#1$}%
    \mathinner{\left\langle\kern-0.8\ht\mstrut\left\langle {#1}\kern0.2\ht\mstrut \big| {#2}\kern0.2\ht\mstrut\right\rangle\kern-0.8\ht\mstrut\right\rangle}%
}

\newcommand{\overlaps}[2]{%
    \sbox{\mstrut}{$#1$}%
    \mathinner{\left\langle\kern-0.9\ht\mstrut\left\langle {#1}\kern0.2\ht\mstrut \big| {#2}\kern0.2\ht\mstrut\right\rangle\kern-0.9\ht\mstrut\right\rangle}%
}

\begin{document}

\title{Variational Quantum Algorithms for Simulation of Lindblad Dynamics}

\author{Tasneem Watad}
 \email{stasnemb@campus.technion.ac.il}
 
\author{Netanel H. Lindner}%
\affiliation{%
 Physics Department, Technion, 3200003 Haifa, Israel}

\date{\today}

\begin{abstract}
We introduce a variational hybrid classical-quantum algorithm to simulate the Lindblad master equation and its adjoint for time-evolving Markovian open quantum systems and quantum observables. Our method is based on a direct representation of density matrices and quantum observables as quantum superstates. We design and optimize low-depth variational quantum circuits that efficiently capture the unitary and non-unitary dynamics of the solutions. We benchmark and test the algorithm on different system sizes, showing its potential for utility with near-future hardware.    
\end{abstract}

\maketitle

\section{\label{sec:intro}Introduction}
Great efforts are currently being put into the development of new generations of quantum computers (QCs), showing impressive progress in terms of their scale and noise rates.
Triggered by this exciting progress, much attention is directed towards unraveling how devices available in the near future can advance various science and technology fields \cite{preskill2018quantum}.
Quantum simulations (QS) \cite{Feynman1982,10.2307/2899535} is considered one of the top natural and immediate applications in which quantum devices can give an advantage over classical digital computers. Numerous quantum algorithms have been proposed for QSs, that aim to approximate the time evolution operator with a controllable error \cite{PhysRevLett.114.090502, berry2007efficient, jordan2012quantum, PhysRevLett.118.010501}. However, these methods are challenging for current and near-term quantum devices which do not operate in the fault-tolerant regime. 
\par
Recent efforts focus on developing a class of variational algorithms based on the approach of hybrid quantum-classical computation which aim to be better suited for near-term, intermediate scale QCs. The most known variational hybrid algorithm is the Variational Quantum Eigensolver (VQE) algorithm \cite{peruzzo2014variational,mcclean2016theory}, which is based on the quantum variational principle and aims to estimate the ground state energy of a physical Hamiltonian. This algorithm regards the quantum device as a co-processor that measures the energy of a quantum system and its derivatives evaluated for a set of variational states, and the optimization of the parameters in the variational manifold is carried by a classical co-processor. Motivated by this algorithm, several variational algorithms for quantum dynamics simulations (VQDS) \cite{PRXQuantum.2.010342, yuan2019theory, li2017efficient, endo2020variational, cirstoiu2020variational, benedetti2021hardware, barison2021efficient,berthusen2021quantum} have been proposed to solve the problem of time-evolving the state of an isolated quantum system. 
Many quantum systems of interest are inevitably coupled to an external environment. In order to study such systems, methods for simulating open systems dynamics are needed. In many cases, a Markovian approximation gives a good description of the dynamics of the external environment. The dynamics of such systems is governed by the Lindblad master equaiton. 

\begin{figure*}
  \centering
  \includegraphics[width=\linewidth, trim={0cm 0cm 0cm 0cm},clip]{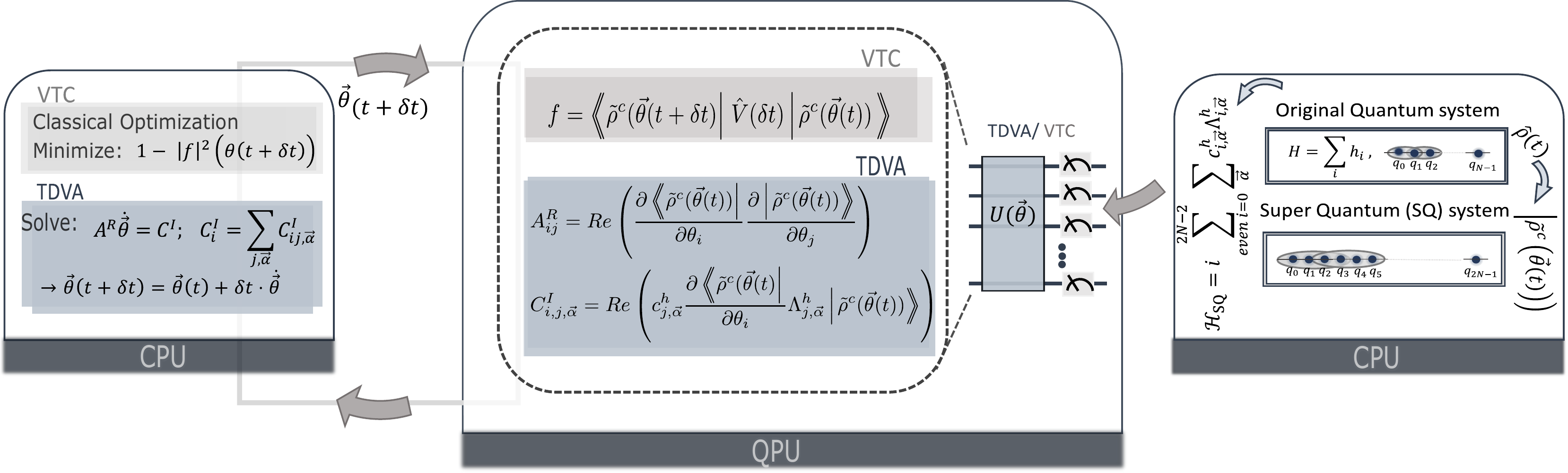}
  \caption{ Schematic illustration of the algorithm for solving the Lindblad master equation. Right: in the first step, the problem of simulating the density matrix operator $\hat{\rho}(t)$ is mapped into time evolving a state $\kettt{\rho(t)}$, living in a larger Hilbert space, under the local Hamiltonian $\mathcal{H}_{SQ}=i\mathcal{L}$, see Eq.~(\ref{eq:linb_supersp}). The graphical illustration here corresponds to the case of a $2$-local Hamiltonian and closed quantum system. Middle and Left: a loop between the classical and quantum co-processors based on either the TDVA or VTC variational algorithms for optimizing the parameters of the ansatz for the evolving state, $\kettt{\tilde{\rho}^c\left(\vec{\theta}(t)\right)}$, defined in Eq.~(\ref{eq:Trotter-like ansatz}). The elements of $A^R$ and $C^I$ defined in Eq.~(\ref{eq:A_C}) and the fidelity $f$, Eq.~(\ref{eq:VTC_fidelity}), needed for TDVA and VTC-based algorithms,  can be computed using a quantum circuit, $U(\vec{\theta})$ of the type described in Appendix \ref{appendix:ansatz construction} that depends on the variational ansatz.}
  \label{fig:sketch_of_the_algorithm} 
\end{figure*}
\par
In this letter, we propose and demonstrate an algorithm for simulating the Lindblad master equation for open quantum systems based on the recently proposed Hybrid variational quantum algorithms, the time-dependent variational algorithm (TDVA) \cite{li2017efficient} and the variational Trotter compression (VTC) algorithm \cite{PRXQuantum.2.010342,berthusen2021quantum,barison2021efficient}. We first present an algorithm for solving the master equation and its adjoint and explain how it can be implemented to time evolve a closed (isolated) quantum system. By generalizing the ansatz used for this task, we obtain a method for performing variational simulations of the Lindblad master equation for open quantum systems with unitary quantum circuits, and without the need for postselection of ancilla qubits \cite{chan2023simulating,
hu2022general,han2021experimental}.
\section{\label{sec:algorithm}Quantum Variational Solver for the Lindblad Master Equation}
The Lindblad equation capturing the dynamics of an open quantum system state $\hat{\rho}$ in the Schroedinger picture is given by \cite{lindblad1976generators,gorini1976completely}
\begin{equation}
\label{eq:lindblad}
    \dot{\hat{\rho}}=\mathcal{L}^s(\hat{\rho})\equiv \mathcal{L}^s_H(\hat{\rho})+\mathcal{D}^s(\hat{\rho}).
\end{equation}
The Liouvillian superoperator $\mathcal{L}^s$ as defined here is composed of a separated unitary part $\mathcal{L}_H^s(\hat{\rho})=-i[\hat{H}, \hat{\rho}]$ and a non-unitary dissipative part, known as the dissipator and is defined as
\begin{equation}
\label{eq:dissip}
\mathcal{D}^s(\hat{\rho})=\sum_{i}{\left[ \hat{L_i}\hat{\rho}\hat{L_i}^{\dagger} -\frac{1}{2}\hat{L_i}^{\dagger}\hat{L_i}\hat{\rho}-\frac{1}{2}\hat{\rho}\hat{L_i}^{\dagger}\hat{L_i} \right]},
\end{equation}
where $\hat{L}_i's$ are the jump (Lindblad) operators of the model. In the following, we use the notation $\kett{\cdot}$ to describe the representation of operators as superstates in the Hilbert–Schmidt space of operators. The representation of the superoperators $\mathcal{L}^s$, $\mathcal{L}_H^s$ and $\mathcal{D}^s$ in this space are denoted as $\mathcal{L}$, $\mathcal{L}_H$ and $\mathcal{D}$, respectively. The evolution equation for the density matrix of $\kettt{\rho(t)}$ is then written as
\begin{equation}
    \label{eq:linb_supersp}
    i\partial_t\kettt{\rho(t)}=i\mathcal{L}\kettt{\rho(t)}\equiv \mathcal{H}_{SQ} \kettt{\rho(t)},
\end{equation}
$\mathcal{H}_{SQ}$ is an effective ``super Hamiltonian" which can be non-Hermitian due to the dissipator part. The formal solution of this equation reads
\begin{equation}
\label{eq:rho_ev}
 \kettt{\rho(t)}=e^{-i\mathcal{H}_{SQ}t}\kettt{\rho(0)}.    
\end{equation}
\par
Analogously, the dynamics of a quantum observable $\hat{O}$ under the Heisenberg pictures is governed by the adjoint master equation,
\begin{equation}
    \label{eq:adj_mstr}
     \dot{\hat{O}}=\mathcal{L}^{s^\dagger}(\hat{O})=\mathcal{L}_H^{s^{\dagger}}+\mathcal{D}^{s^{\dagger}}(\hat{O}),
\end{equation}
$\mathcal{L}^{s^{\dagger}}$ is known as the adjoint Liouvillian, and we introduce the adjoint super Hamiltonian defined as $\mathcal{H}_{SQ}^\dagger \coloneqq i\mathcal{L}^\dagger$ where $\mathcal{L}^\dagger$ is the operator representation of $\mathcal{L}^{s^\dagger}$. 
Similar to the density matrix, regarding the quantum observable $\hat{O}$ as a superstate, the evolution of $\hat{O}$ will then be given by $\kettt{O(t)}=\exp\left(-i\mathcal{H}_{SQ}^{\dagger}t\right)\kettt{O(0)}$.
Throughout this section, we will focus on solving Eq.~(\ref{eq:lindblad}). However, the method can be straightforwardly adapted for obtaining a variational solution for $\kettt{O(t)}$.
\par
A central step of the algorithm is mapping the time evolution problem of the density matrix $\kett{\rho}$, representing the state of a quantum system of $N$ qubits which we will refer to as ``spins", (or similarly, the quantum operator $\ketttt{O}$), into a problem of time evolving a quantum system of $2N$ qubits state under a local Hamiltonian $\mathcal{H}_{SQ}$ ($\mathcal{H}_{SQ}^{\dagger}$).
To achieve this, we represent the superstate $\kett{\rho}$ on a quantum computer comprising of $2N$ qubits using the identification: $\kett{\rho}=\sum_{\vec{\alpha}} c_{\vec{\alpha}}$ $\ketttt{P_{\vec{\alpha}}},$
where ${\vec{\alpha}}=(\alpha_0,\alpha_1,\cdots,\alpha_N)$, $\alpha_j\in\{0,x,y,z\}$, $c_{\vec{\alpha}}$'s are real coefficients and $\ketttt{P_{\vec{\alpha}}}$’s are the superstate representation of the regular Pauli basis elements $P_{\vec{\alpha}}$'s. The $P_{\vec{\alpha}}$'s are represented via the map
\begin{equation}
    P_{\vec{\alpha}}=\frac{1}{2^{N/2}} \otimes_{j=0}^{N-1} \sigma_{j}^{\alpha_j} \rightarrow \ketttt{P_{\vec{\alpha}}}= \frac{1}{2^{N/2}} \otimes_{j=0}^{N-1} \ketttt{\sigma_{j}^{\alpha_j}}
\end{equation}
and $\frac{1}{\sqrt{2}}\ketttt{\sigma_{j}^{\alpha_j}}$ is equal to $\ketttt{00}$, $\ketttt{01}$, $\ketttt{10}$ and $\ketttt{11}$ for $\alpha_j=0,x, y$ and $z$ respectively.
In the new basis, we consider $k_1$-local jump operators $L_j$'s, see Eq.~(\ref{eq:dissip}), and $k_2$-local Hamiltonian of the form $H=\sum_j{h_j}$. i.e., each $L_j$ and $h_j$ acts on at most $k_1$ and $k_2$ qubits respectively.   Under these locality conditions, $\mathcal{L}$ will be $2\cdot max(k_1,k_2)-$local. In addition, each unit cell composed of one spin in the original physical system is mapped into a larger unit cell composed of two qubits.
For the single-site jump operators and $2$-local Hamiltonian case, i.e., $k_1=1$ and $k_2=2$, $\mathcal{L}_H$ and $\mathcal{D}$ read as
\begin{equation}
\label{eq:op_liouv}
\begin{aligned}
    &\mathcal{L_H}=\sum_{\underset{even}{i=0}}^{2N-2}\sum_{\vec{\alpha}}{c^h_{i,{\vec{\alpha}}} \Lambda^h_{i,{\vec{\alpha}}}}, \\
    &\mathcal{D}=\sum_{\underset{even}{i=0}}^{2N-2}\sum_{\vec{\beta}}{c^u_{i,{\vec{\beta}}} \Delta^u_{i,{\vec{\beta}}}}+\sum_{\underset{even}{i=0}}^{2N-2}\sum_{\vec{\gamma}}{c^n_{i,{\vec{\gamma}}} \Delta^n_{i,{\vec{\gamma}}}} 
\end{aligned}
\end{equation}
where $\vec{\alpha}=(\alpha_0,\alpha_1,\alpha_2,\alpha_3)$, $\vec{\beta}=(\beta_0,\beta_1)$, $\vec{\gamma}=(\gamma_0,\gamma_1)$ and $\alpha_i$, $\beta_i$ and $\gamma_i$ all take values $\in \{ 0,x,y,z \}$. 
Here $\Lambda^h_{i,{\vec{\alpha}}}=\sigma^{\alpha_0}_i\sigma^{\alpha_1}_{i+1}\sigma^{\alpha_2}_{i+2}\sigma^{\alpha_3}_{i+3}$ and $c^h_{i,{\vec{\alpha}}}$ are pure imaginary coefficients. Note that when the dissipator, $\mathcal{D}^s$, is mapped to an operator acting on the superstate, it can be decomposed into its Hermitian and an anti-Hermitian components. These components correspond to $\Delta^n_{i,{\vec{\gamma}}}=\sigma^{\gamma_0}_i\sigma^{\gamma_1}_{i+1}$ and $\Delta^u_{i,{\vec{\beta}}}=\sigma^{\beta_0}_i\sigma^{\beta_1}_{i+1}$ respectively, where $c^n_{i,{\vec{\gamma}}}$'s are real coefficients and $c^u_{i,{\vec{\beta}}}$'s are purely imaginary.
Given this representation, we use the Hamiltonian $\mathcal{H}_{SQ}$ to time evolve the superstate $\kett{\rho}$ using a variational algorithm for time-evolving quantum states.
To time evolve the superstate $\kett{\rho}$ under the super Hamiltonian $\mathcal{H}_{SQ}$, we consider a family of parametrized superstates, $\ketttt{\tilde{\rho}\big(\vec{\theta}(t)\big)},$ which comprise a variational manifold in the Hilbert-Schmidt space. Our goal is to find the parameters $\vec{\theta}$ for which $\ketttt{\tilde{\rho}\big(\vec{\theta}(t)\big)}$ gives the best description of $\kettt{\rho(t)}$. To this end, we will consider two types of hybrid classical-quantum variational algorithms: the time-dependent variational algorithm (TDVA) \cite{li2017efficient} and the variational Trotter compression (VTC) algorithm \cite{PRXQuantum.2.010342,berthusen2021quantum,barison2021efficient}. The TDVA method based on the classical McLachlan’s variational principle \cite{mclachlan1964variational} is obtained from minimizing the error of the evolution of the variational superstate $\ketttt{\tilde{\rho}\big(\vec{\theta}(t)\big)},$ according to the evolution equation Eq.~(\ref{eq:linb_supersp}), i.e., minimizing the cost-function
\begin{equation}
    L^2=\norm{\left( d/dt+i\mathcal{H}_{SQ} \right)\kettt{\tilde{\rho}\left(\vec{\theta}\left(t\right)\right)}}^2.
\end{equation}
Assuming $\vec{\theta}$ is real, this minimization yields a set of linear equations of motion for the variational parameters $\vec{\theta}$, given by
\begin{equation}
    \sum_j A_{ij}^R\dot{\theta_j}=C_i^I, 
    \label{eq:algebr_eq}
\end{equation}
where the elements of the real part of matrix $A^R$ and the imaginary part of vector $C^I$ are 
\begin{equation}
\label{eq:A_C}
\begin{aligned}
&A_{ij}^R= Re\left( \frac{\partial \bbra{\tilde{\rho}\big(\vec{\theta}(t)\big)}}{\partial\theta_i} \frac{\partial \kett{\tilde{\rho}\big(\vec{\theta}(t)\big)}}{\partial\theta_j} \right)\\
& C_i^I=Im\left( \frac{\partial \bbra{\tilde{\rho}\big(\vec{\theta}(t)\big)}}{\partial\theta_i}\mathcal{H}_{SQ}\kett{\tilde{\rho}\big(\vec{\theta}(t)\big)}\right).
\end{aligned}
\end{equation}
While Eq.~(\ref{eq:A_C}) was originally considered for simulations of a quantum state evolving under a Hermitian Hamiltonian, in Appendix \ref{appendix:deriv_mach} we show that it holds also in the case of a non-Hermitian $\mathcal{H}_{SQ}$.
Due to the Hermitian part of the dissipator, Eq.~(\ref{eq:op_liouv}), the state $\kettt{\tilde{\rho}(t)}$ is not normalized. However, as we will show in the following section, we can rewrite this ansatz as 
\begin{equation}
\label{eq:ansatz_normalization1}
    \ketttt{\tilde{\rho}\big(\vec{\theta}(t)\big)}=a(t)\cdot \mathcal{R}_m\cdots \mathcal{R}_2\cdot \mathcal{R}_1 \kettt{0^{\otimes 2N}}\equiv a\ketttt{\tilde{\rho}^u\big(\vec{\theta}(t)\big)},
\end{equation}
where the gates ${\mathcal{R}_m\cdots \mathcal{R}_2 , \mathcal{R}_1}$ are unitaries and depend on the parameters $\vec{\theta}$, $a(t)$ is a complex parameter and $\ketttt{\tilde{\rho}^u}$ is a normalized state. 
With this form, and regarding $a(t)$ as a complex variational parameter, the elements of $A^R$ and $C^I$ in  Eq.~(\ref{eq:algebr_eq}) can be expressed as a linear combination of expectation values of the form

\begin{equation}
\begin{aligned}
    \bbbra{0^{\otimes 2N }}&\left(\otimes_{k=1}^m \mathcal{W}^\dagger_{m-k+1}\mathcal{R}_{m-k+1}^\dagger\right)\cdot\\
    &U_{\mathcal{L}}\left(\otimes_{k=1}^m \mathcal{R}_{k}\mathcal{V}_k\right) \kettt{0^{\otimes 2N}}.
\end{aligned}
\end{equation}

The gates $\mathcal{V}_k$'s and $\mathcal{W}^\dagger_k$'s are obtained from the derivatives of $\mathcal{R}_k$'s and their complex conjugates, respectively, and are trivial for most values of $k$'s.  $\mathcal{U}_{\mathcal{L}}$ is either trivial for terms corresponding to $A^R$ elements or it belongs to the set $\{\Lambda_{i,\vec{\alpha}}^h, \Delta_{i,\vec{\beta}}^u, \Delta_{i,\vec{\gamma}}^n\}$'s for terms corresponding to $C^I$ elements. The variational algorithm proceeds as follows. First, the elements $A^R_{ij}(t)$ and  $C^I_i(t)$ are evaluated on the quantum computer. Appendix \ref{appendix:ansatz construction} explains how to evaluate these elements using a quantum computer with short-depth circuits for unitary $\mathcal{R}_k$'s. We also demonstrate how similar circuits can be employed for a special ansatz involving non-unitary gates, which will be introduced in the next section. The number of measurements needed to evaluate these terms can be remarkably decreased using schemes based on the recently proposed classical shadows protocol \cite{huang2020predicting, nakaji2022measurement}. In the second step, the results are fed to a classical computer, which inverts the algebraic linear equations, Eq.~(\ref{eq:algebr_eq}), and finds $\dot{\theta_j}$'s from which we obtain the trial superstate at time $t+\delta t$, given by
$\kettt{\tilde{\rho}\big(\vec{\theta}(t)+\delta t\cdot \dot{\vec{\theta}}\big)}$ where $\delta t$, the evolution time-step, is small. 
\par
For simulations based on the VTC method, we consider the ansatz described in Eq.~(\ref{eq:ansatz_normalization1}).
The variational cost function is then defined as $1-|f|^2(t)$, with
\begin{equation}
\label{eq:VTC_fidelity}
|f|^2(t) = \biggl|\frac{a(t-\delta t)}{a(t)} \bbra{\tilde{\rho}^u\big(\vec{\theta}(t)\big)}\hat{V}\left(\delta t\right) \kett{\tilde{\rho}^u\big(\vec{\theta}(t-\delta t)\big)}\biggl|^2.
\end{equation} 
The variational parameters are then optimized to minimize this cost function. To achieve this, since $a(t)$ cannot be considered as a variational parameter, an alternative approach is required to either measure it or express it in terms of the parameter set $\vec{\theta}(t)$. One approach for measuring $a(t)$ at each time step is through the time evolution of the state's purity.
The purity of the state, $\mathcal{P}=Tr\big(\tilde{\rho}^2\big)=|a|^2$, evolves in time according to
\begin{equation}
    \frac{d\mathcal{P}}{dt} = Tr\biggl(\tilde{\rho}\frac{d\tilde{\rho}}{dt}+\frac{d\tilde{\rho}}{dt}\tilde{\rho}\biggl)
\end{equation}
Substituting Eq.~(\ref{eq:lindblad}) and the approximation $d\mathcal{P}/dt \approx \big(|a(t)|^2-|a(t-\delta t)|^2)/\delta t$, and rearranging this equation gives 
\begin{equation}
\label{eq:a_evol}
    \biggl|\frac{a(t-\delta t)}{a(t)}\biggl|=\biggl(1+2\delta t\bbra{\tilde{\rho}^u\big(\vec{\theta}(t-\delta t)\big)}\mathcal{D}\left(\delta t\right) \kett{\tilde{\rho}^u\big(\vec{\theta}(t-\delta t)\big)}\biggl)^{-1/2}.
\end{equation}
$\mathcal{D}$ is the dissipator part of the Liouvillian, Eq.~(\ref{eq:op_liouv}).
By measuring this expression at each time step, the fidelity, Eq.~(\ref{eq:VTC_fidelity}), can be expressed using only the set of variational parameters $\vec{\theta}(t)$ that needs to be optimized.

An alternative method for handling the unknown purity, is by imposing the trace preserving property. This property gives the equation $\overlapsop{P_0}{\tilde{\rho}\big(\vec{\theta}(t)\big)}=\frac{1}{2^{N/2}}$ where $\ketttt{P_0}=\ketttt{00\cdots 00}$ corresponds to $P_0=\frac{1}{2^{N/2}} I^{\otimes N}$, yielding
\begin{equation}
    \label{eq:a_factor}
    a=\left(2^{N/2}\overlaps{P_0}{\tilde{\rho}^u}\right)^{-1}.
\end{equation}
Again, substituting Eq.~(\ref{eq:a_factor}) in Eq.~(\ref{eq:VTC_fidelity}), reduces the optimization problem to optimizing over $\vec{\theta}(t)$ which are the only unknown variables.
As long as the steady state purity $\mathcal{P}$ is not of $O(1)$, indicating that the state is not pure or near-pure, the measured overlap between $P_0$ and $\tilde{\rho}^u$, denoted by $\overlaps{P_0}{\tilde{\rho}^u}$, is not expected to be exponentially small in the number of physical spins, nor is it expected that the error obtained from measuring $a(t)$ will be large. If this condition is not met, extracting $a$ using Eq.~(\ref{eq:a_evol}) is the preferred method.

To address the non-unitarity problem associated with $V(\delta t)$, one can decompose it as 
\begin{equation}
    V(\delta t)=I^{\otimes 2N}+\mathcal{L}\delta t+\mathcal{L}^2\delta t^2+\cdots \mathcal{L}^n\delta t^n+ O(\delta t^{n+1}).
\end{equation}
Using this decomposition, one can rewrite the fidelity as a sum of expectation values of unitaries.
The value of $n$ can be selected to ensure a reasonable number of measurements is obtained while also having a good approximation for $V(\delta t)$.
The optimal variational parameters, $\vec{\theta}(t)$ can be determined by classically minimizing the cost function $1-|f|^2$ (refer to Appendix \ref{appendix:ansatz construction} for different methods to measure this quantity).

A schematic diagram of the algorithm demonstrated on time evolving a closed system is presented in Fig.~\ref{fig:sketch_of_the_algorithm}.
\par
\section{\label{sec:ansatz_construction}Ansatz Construction}
Embedding the physical properties of the specific problem into the design of the ansatz is crucial for optimizing the algorithm's performance. We focus on variational ansatze inspired by the conventional Suzuki-Trotter approximation structure.
\subsection{\label{ Ansatz_for_unitary_time_evolution} Ansatz for unitary time evolution} We first consider the case of a closed quantum system. For a quantum state $\kett{\rho}$ that time evolves under Liouvillian $\mathcal{L}=\mathcal{L}_H$ as described in Eq.~(\ref{eq:op_liouv}), the unitary time evolution superoperator $\mathcal{U}(t)=\exp\left(-i\mathcal{H}_{SQ} t\right)$ can be approximated using the Suzuki-Trotter expansion as 
\begin{equation}
\label{eq:Trottereization_eq}
    \mathcal{U}(t)=e^{-i\mathcal{H}_{SQ}t}=\left( \Pi_{i:even,\vec{\alpha}} e^{c^h_{i,{\vec{\alpha}}} \Lambda^h_{i,{\vec{\alpha}}} t/n} \right)^n+O(t^2/n).
\end{equation}
Here, $\Delta t=t/n$ is small, and each operation $\hat{V}(t)=\Pi_{i:even,\vec{\alpha}} \ \exp\left(c^h_{i,{\vec{\alpha}}} \Lambda^h_{i,{\vec{\alpha}}} t/n\right)$ is a single Trotter step.
The corresponding variational ansatz takes the form
\begin{equation}
\label{eq:Trotter-like ansatz}
\kett{\tilde{\rho}^c\big(\vec{\theta} (t)\big)}= \Pi_{r=1}^m \Pi_{i:\text{even},\vec{\alpha}} {\mathcal{U}_{i,\vec{\alpha}}^h \left( \theta_{i,r}^{\vec{\alpha}}\right)}\ketttt{\rho(0)},
\end{equation}
where $\vec{\theta} =\{\theta_{i,r}^{\vec{\alpha}}\}_{i,r,\vec{\alpha}}$ and $\mathcal{U}_{i,\vec{\alpha}}^h\left(\theta_{i,r}^{\vec{\alpha}}\right) =\exp\left(\theta_{i,r}^{\vec{\alpha}} c^h_{i,\vec{\alpha}} \Lambda^h_{i,\vec{\alpha}}\right)$.
Equivalently, the ansatze are composed of a sequence of ``large'' Trotter steps. In this ansatz, the increment $\Delta t$ in the conventional Trotterization approximation is replaced by $\theta_{i,r}^{\vec{\alpha}}$ that can take any value in the range $[0,2\pi)$. This type of ansatz gives an efficient description of the superstate $\kettt{\rho(t)}$ if it gives a good approximation with a small value of $m$.
\par
In the limit of short evolution time, the exact evolved state resides on the tangent space of the Trotter ansatz's variational manifold, providing an accurate description of the state for small values of $t$. This is crucial for the algorithm's performance in approximating the evolution at large times, as each time step heavily relies on the evolution history.
\par
Moreover, the unitary operator $\exp\left(\theta_{i,r}^{\vec{\alpha}} c^h_{i,{\vec{\alpha}}} \Lambda^h_{i,{\vec{\alpha}}}\right)$ representing a four-qubit or three-qubit gate can be efficiently decomposed into single-qubit and two-qubit gates, as demonstrated in \cite{clinton2021hamiltonian}. Additionally, any two-qubit gate can be decomposed into single-qubit gates and CNOT gates, as shown in \cite{PhysRevA.69.032315, PhysRevA.69.010301, PhysRevA.63.062309}.

We further partition the set of gates $\{\mathcal{U}_{i,\vec{\alpha}}^h \}_{i,\vec{\alpha}}$ into completely commuting subsets, where the chosen partition corresponds to the fewest number of subsets. Within each Trotter step, all the gates, $\{\mathcal{U}_{i,\vec{\alpha}}^h \}_{i,\vec{\alpha}}$, within each subset, are assigned an equal parameter. This significantly reduces the parameters vector $\vec{\theta}$. The resulting ansatz is denoted by $\kett{\tilde{\rho}^c_1(\vec{\theta})}$ (see Appendix~\ref{appendix:ansatz construction} for details about the construction of the anzats).

\subsection{\label{sec:Ansatz_Lindb}Ansatz for Lindblad dynamics}
In the open system problem, since the dynamics of $\hat{\rho}$ under Eq.~(\ref{eq:lindblad}) is non-unitary, the norm of the superstate $\kettt{\rho(t)}$, $\sqrt{\normsop{\rho(t)}}$, changes throughout the evolution. 
A trotter-like ansatz for $\kettt{\rho(t)}$ is given by
\begin{equation}
\label{eq:open_system_trotter_ansatz_start}
\begin{aligned}
    &\kett{\tilde{\rho}(\vec{\theta})}=\Pi_{r=1}^m\biggl[ \biggl(\Pi_{i:even}\Pi_{\vec{\alpha}}{\mathcal{U}_{i,\vec{\alpha}}^h \left( \theta_{i,r}^{\vec{\alpha}}\right)}\biggr)\cdot\\&\biggl(\Pi_{i:even}\Pi_{\vec{\beta}}{\mathcal{U}_{i,{\vec{\beta}}}^u \left( \theta_{i,r}^{\vec{{\beta}}}\right)}\biggr)\cdot \biggl(\Pi_{i:even}\Pi_{\vec{\gamma}}{\mathcal{G}_{i,\vec{\gamma}}^n \left(\theta_{i,r}^{\vec{\gamma}}\right)}\biggr)\biggr]\kettt{\rho(0)}
\end{aligned}
\end{equation}
Where $\vec{\theta} =\{\{\theta_{i,r}^{\vec{\alpha}}\}_{i,r,\vec{\alpha}}, \{\theta_{i,r}^{\vec{\beta}}\}_{i,r,\vec{\beta}}, \{\theta_{i,r}^{\vec{\gamma}}\}_{i,r,\vec{\gamma}}\}$. $\mathcal{G}_{i,\vec{\gamma}}^n\left(\theta_{i,r}^{\vec{\gamma}}\right)=\exp\left(\theta_{i,r}^{\vec{\gamma}}c^n_{i,{\vec{\gamma}}} \Delta^n_{i,{\vec{\gamma}}}\right)$ are non-unitaries corresponding to the non-unitary part of the dissipator $\mathcal{D}$, and $\mathcal{U}_{i,{\vec{\beta}}}^u\left( \theta_{i,r}^{\vec{\beta}}\right)=\exp\left(\theta_{i,r}^{\vec{\beta}}c^u_{i,{\vec{\beta}}} \Delta^u_{i,{\vec{\beta}}}\right)$ are unitaries corresponding to the unitary part of the dissipator $\mathcal{D}$. As aforementioned, $\mathcal{U}_{i,\vec{\alpha}}^h\left( \theta_{i,r}^{\vec{\alpha}}\right) =\exp \left(\theta_{i,r}^{\vec{\alpha}} c^h_{i,{\vec{\alpha}}} \Lambda^h_{i,{\vec{\alpha}}}\right)$,  corresponding to $\mathcal{L}_H$. See Eq.~(\ref{eq:op_liouv}) for the decomposition of $\mathcal{L}_H$ and $\mathcal{D}$. Similar to $\kettt{\tilde{\rho}_1^c}$, all gates corresponding to $\mathcal{L}_H$ involved in each Trotter step will be partitioned into commuting subsets, and equal parameters are assigned to all the variational gates within each subset. A similar partition is performed for the all the gates corresponding to the unitary part of the dissipation. 
\par
 In the context of non-unitary dynamics, a significant challenge lies in efficiently implementing the non-unitary gates,  $\mathcal{G}_{i,\vec{\gamma}}^n$'s, involved in the quantum circuits needed for computing matrices $A$ and $C$ in the TDVA framework and the fidelity $f$ in the VTC framework. One approach to achieve this is to embed the non-unitary gate in a larger unitary gate acting on an ancilla qubit, utilizing post-selection or projection \cite{ daskin2017ancilla, terashima2005nonunitary,PRXQuantum.2.010342}. However, as the number of such non-unitary gates is linear in the system size, the postselection procedure will result in an exponential cost. To avoid this cost, we propose an alternative method of replacing the dissipative non-unitary part with a unitary implementation. This is based on the fact that given a non-unitary gate $G$ acting on a given qubits quantum state $|\psi\rangle$, it is always possible to find a unitary $U$ such that 
\begin{equation}
\label{eq:unit-non-unit}
    G|\psi\rangle=a\cdot U|\psi\rangle,
\end{equation}
where $a$ is a complex factor. Note that $U$ and $a$ depend on $G$ and $|\psi\rangle$. 
If $\kettt{\rho(0)}$ is a product state, and the jump operators $\hat{L}_i$'s in Eq.~(\ref{eq:dissip}) are single-site, a simplification can be made for the first large Trotter step of the non-unitary part. Namely, a set of 2-qubit unitaries $\mathcal{U}_i^n$'s, simulating the action of the non-unitary gates $\Pi_{\vec{\gamma}}\mathcal{G}_{i,\vec{\gamma}}^n(\theta_{i,r=1}^{ \vec{\gamma}})$'s according to Eq.~(\ref{eq:unit-non-unit}) can be classically computed. Each unitary $\mathcal{U}_i^n$ satisfies $a_{i,1}\mathcal{U}_i^n \kettt{\rho(0)} =  \Pi_{\vec{\gamma}}\mathcal{G}_{i,\vec{\gamma}}^n(\theta_{i,r=1}^{ \vec{\gamma}}) \kettt{\rho(0)}$, where $a_{i,1}$ is a complex factor. In a similar way, a set of unitaries $\mathcal{U}_{i,\vec{\gamma}}^n$'s for simulating the derivative of $\Pi_{\vec{\gamma}}\mathcal{G}_{i,\vec{\gamma}}^n(\theta_{i,r=1}^{ \vec{\gamma}})$'s with respect to 
$\theta_{i,r=1}^{ \vec{\gamma}}$'s and the corresponding complex factors $a_{i, r=1,\vec{\gamma}}$'s can be classically computed. 
However, this is more challenging for the later Trotter steps $(r>1)$ where the computation qubits are already in an entangled state. For these Trotter steps, we replace each one of the terms $\Pi_{\vec{\gamma}}\mathcal{G}_{i,\vec{\gamma}}^n(\theta_{i,r}^{\vec{\gamma}})$ in the ansatz, Eq.~(\ref{eq:open_system_trotter_ansatz_start}), with $a_{i,r}\mathcal{G}^g_i$, where $a_{i,r}$ is a complex factor and $\mathcal{G}^g_{i}$ is a general two-qubit variational quantum circuit. Each of these circuits depends on $15$ variational parameters: $\vec{\theta}^g_{r,i} = \left(\theta_{r,i,0}^g, \theta_{r,i,1}^g\cdots \theta_{r,i, 14}^g\right)$. The obtained ansatz is denoted by $\kett{\tilde{\rho}^o(\vec{\theta})}$. The set $\{\theta_{i,r=1}^{\vec{\gamma}}\}_{i,\vec{\gamma}}$ is reduced to $\{\theta_{i,r=1}^{\vec{\gamma}}\}^\prime_{i,\vec{\gamma}}$ using a similar partition to this performed on the unitary parts.
This ansatz can be rewritten as
\begin{equation}
    \label{eq:ansatz_normalization}
    \kettt{\tilde{\rho}^o\left(\vec{\theta}(t)\right)}=a(t)\kettt{\tilde{\rho}^u\left(\vec{\theta}(t)\right)}
\end{equation}
where $\ketttt{\tilde{\rho}^u}$ is a quantum state with a unit norm that is given by unitaries acting on the initial state,  and $a(t)=\Pi_{i,r>0} a_{i,r}$.
\par 
We use this ansatz, $\kettt{\tilde{\rho}^o\left(\vec{\theta}(t)\right)}$ for both the TDVA and VTC methods. With this form, the elements of $A^R$ and $C^I$, Eq.~(\ref{eq:algebr_eq}), and the fidelity $f$, Eq.~(\ref{eq:VTC_fidelity}), can be rewritten as a combination of expectation values of unitaries, and thus can be evaluated on a quantum computer using short depth circuits as shown in Appendix \ref{appendix:ansatz construction}.  
\section{\label{sec:NE}Numerical experiments}
\subsection{\label{sec:NE-Closed}Application to dynamics of quantum observables of a closed system}
To test the performance of the algorithm and the choice of the variational ansatz we introduced in the previous section, we simulate it numerically for the closed transverse field Ising model described by the Hamiltonian
\begin{equation}
\label{eq:TFI_equation}
    H=\sum_{i=0}^{N-1} \left( J\sigma_i^z \sigma_{i+1}^z+h\sigma_i^x \right),
\end{equation}
with closed boundary conditions, $\sigma^z_N=\sigma^z_0$ and $J=h$. We study the performance of the variational algorithm by computing the dynamics of $O(t=0)=\sigma_1^y$. We use $O(t)$ and $\tilde{O}\big(\vec{\theta}(t)\big)$ to denote the exact and variational solutions of the time evolution problem. 
\begin{figure*}[]
  \centering
    \includegraphics[width=0.9\linewidth, trim={0cm 1cm 0cm 0cm},clip,scale=0.7]{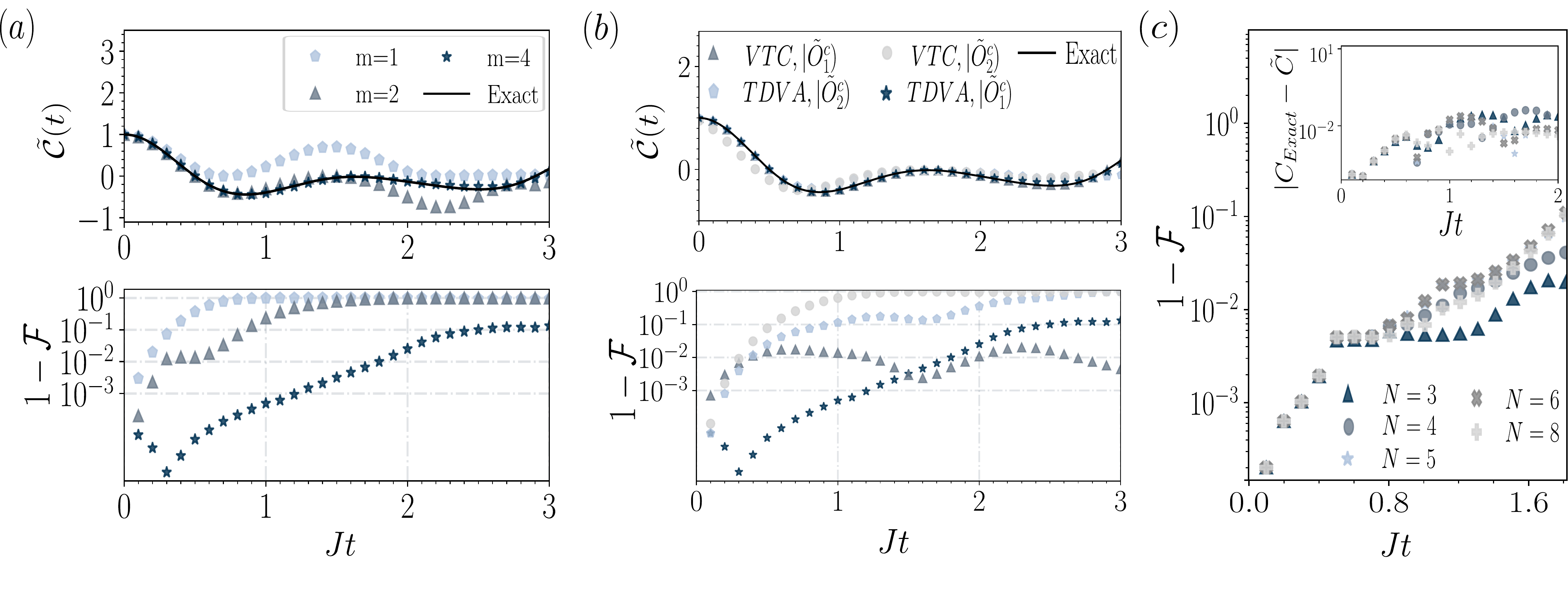}
  \caption{Numerical simulations of variational quantum algorithms for approximating Heisenberg evolution of operators. The operator $O(0)=\sigma_1^y$ is evolved according to the transverse field Ising model Hamiltonian, Eq.~(\ref{eq:TFI_equation}) with J=h. (a) Upper panel: the auto-correlation function, $\tilde{\mathcal{C}}(t)=\overlapsop{O(0)}{\tilde{O}(t)}$, simulated using the variational ansatz $\kett{\tilde{O}(t)}=\kett{\tilde{O}_1^c(\vec{\theta}(t))}$, of a similar structure as $\kett{\tilde{\rho}^c_1(\vec{\theta})}$ described in Eq.~(\ref{eq:Trotter-like ansatz}), for different values of $m$ corresponding to the number of large Trotter steps. The TFI model here is defined for N=3 spins. The exact dynamics is plotted as a solid black curve.
  Lower panel: the infidelity, $1-\mathcal{F}$, with $\mathcal{F}=\biggl|\overlapsop{O(t)}{\tilde{O}_1^c\big(\vec{\theta}(t)\big)}\biggr|^2$. (b) Same as (a), but here showing a comparison between the performance of the TDVA-based algorithm and the VTC-based algorithm for the two variational ansatze $\kett{\tilde{O}_1^c(\vec{\theta}(t)}$ and $\kett{\tilde{O}_2^c(\vec{\theta}(t))}$ defined in Section \ref{sec:NE-Closed}, for N=3. (c) Performance for different system sizes (number of spins), for the same setting as in (a). We plot the infidelity for the ansatz $\kett{\tilde{O}_1^c}$ with $m=2$ for short times and $m=4$ for $Jt\geq0.5$. Inset: absolute error $\big|\mathcal{C}_{Exact}-\tilde{\mathcal{C}}\big|$, as a function of time, where $\mathcal{C}_{Exact}$ corresponds to the exact value. 
  Tikhonov regularization method was implemented here to tackle the rank deficiency issue of the $A^R$ matrix in Eq.~(\ref{eq:algebr_eq}) (see Appendix \ref{appendix:tikhonov}).}
  \label{fig:First_and_Second_Figures_Combined3}
\end{figure*}
Fig.~\ref{fig:First_and_Second_Figures_Combined3}(a) shows a comparison of the performance of the algorithm employing the family of trial states $\kett{\tilde{O}(\vec{\theta})}=\kett{\tilde{O}_1^c(\theta)}$ which has the same structure as $\kett{\tilde{\rho}^c_1(\vec{\theta})}$ described in Eq.~(\ref{eq:Trotter-like ansatz}) where now $\mathcal{U}_{i,\vec{\alpha}}^h\left( \theta_{i,r}^{\vec{\alpha}}\right) =\exp\left(\theta_{i,r}^{\vec{\alpha}} (c^h_{i,{\vec{\alpha}}} \Lambda^h_{i,{\vec{\alpha}}})^\dagger\right)$. The figure shows a comparison for different values of $m$, i.e., different numbers of large Trotter layers composing the ansatz. Here, $N=3$, and the time evolution is performed using the TDVA algorithm. The quantity presented in the upper panel figure is the auto-correlation function $\tilde{\mathcal{C}}(t)=\overlapsop{O(0)}{\tilde{O}(t)}$. This quantity can be computed on a QC given the time evolved operator, $\kett{\tilde{O}(t)}$ using the same methods described in the previous section for evaluating $|f|^2$ for the VTC approach. In the lower panel of Fig.~\ref{fig:First_and_Second_Figures_Combined3}(a) we show the infidelity defined as $1-\mathcal{F}=1-\bigg|\overlapsop{O(t)}{\tilde{O}_1^c\big(\vec{\theta}(t)\big)}\bigg|^2$. Increasing $m$ leads to a better approximation of the superstate, and this improvement becomes more significant as the simulation time increases.
\par
Next, we turn to compare the efficiency of the family of ansatze we introduced previously $\kett{\tilde{O}_1^c\big(\vec{\theta}(t)\big)}$ with another family that we call, $\kett{\tilde{O}_2^c(\vec{\theta})}$. The family of ansatze $\kett{\tilde{O}_2^c(\vec{\theta})}$ is of a similar structure as $\kett{\tilde{O}_1^c(\vec{\theta})}$ but with the difference that all the large Trotter steps are identical, i.e., $\theta_{i,r}^{\vec{\alpha}}$'s are equal for different $r$'s. Fig.~\ref{fig:First_and_Second_Figures_Combined3}(b) shows a comparison between employing those two ansatze for $m=4$, again for $N=3$, where the time-evolution is performed based on both the VTC algorithm based on Sequential Least Squares Programming (SLSQP) optimization method and TDVA.
 As it can be clearly seen from the lower panel of Fig.~\ref{fig:First_and_Second_Figures_Combined3}(b), showing the infidelity $1-\mathcal{F}$, $\kett{\tilde{O}_1^c(\vec{\theta})}$ outperforms $\kett{\tilde{O}_2^c(\vec{\theta})}$ in representing the exact superstate evolution. In addition, the TDVA-based algorithm and the VTC-based algorithm both effectively capture the system dynamics. We finally examine the scaling behavior of the algorithm and the large Trotter steps ansatz efficiency with system size. 
 Fig.~\ref{fig:First_and_Second_Figures_Combined3}(c) shows the infidelity and the absolute error, which is defined as the difference between the exact auto-correlation function $\mathcal{C}_{Exact}(t)=\overlapso{O(0)}{O(t)}$ and the value obtained from the algorithm $\tilde{\mathcal{C}}(t)$, for different system sizes. Here, for the beginning of the evolution, a 2-large Trotter step ansatz, of the form $\kett{\tilde{O}^c_1}$, is employed, whereas for the remaining time evolution, a 4-large Trotter step ansatz is used. Overall, the algorithm using the ansatz of the form $\kett{\tilde{O}^c_1}$, continues to perform well for larger system sizes. Notably, we observe that the evaluation of the auto-correlation function, as shown in the inset, does not deteriorate when increasing system size. We expect, though, that at large times, the number of layers needed to maintain a fidelity below some threshold scales exponentially with system size, as shown in \cite{PhysRevResearch.4.023097}. 
 \par
When simulating the dynamics 
with the TDVA-based algorithm, the matrix $A^R$ can become singular and rank-deficient. In particular, this occurs for relatively complex ansatze. These issues hamper inverting the algebraic linear equations, Eq.~(\ref{eq:algebr_eq}), and consequently severely affects the accuracy of the results even when applying least-squares methods. To address these issues and prevent them from reducing the performance of the algorithm, we implemented the Tikhonov regularization method when solving Eq.~(\ref{eq:algebr_eq}), as suggested in \cite{yuan2019theory} as explained in detail in Appendix \ref{appendix:tikhonov}.

\par
\subsection{\label{sec:NE-open}Application to dynamics of open quantum system}
\begin{figure}[]
  \includegraphics[width=\linewidth, trim={0cm 0cm 0cm 0cm},clip]{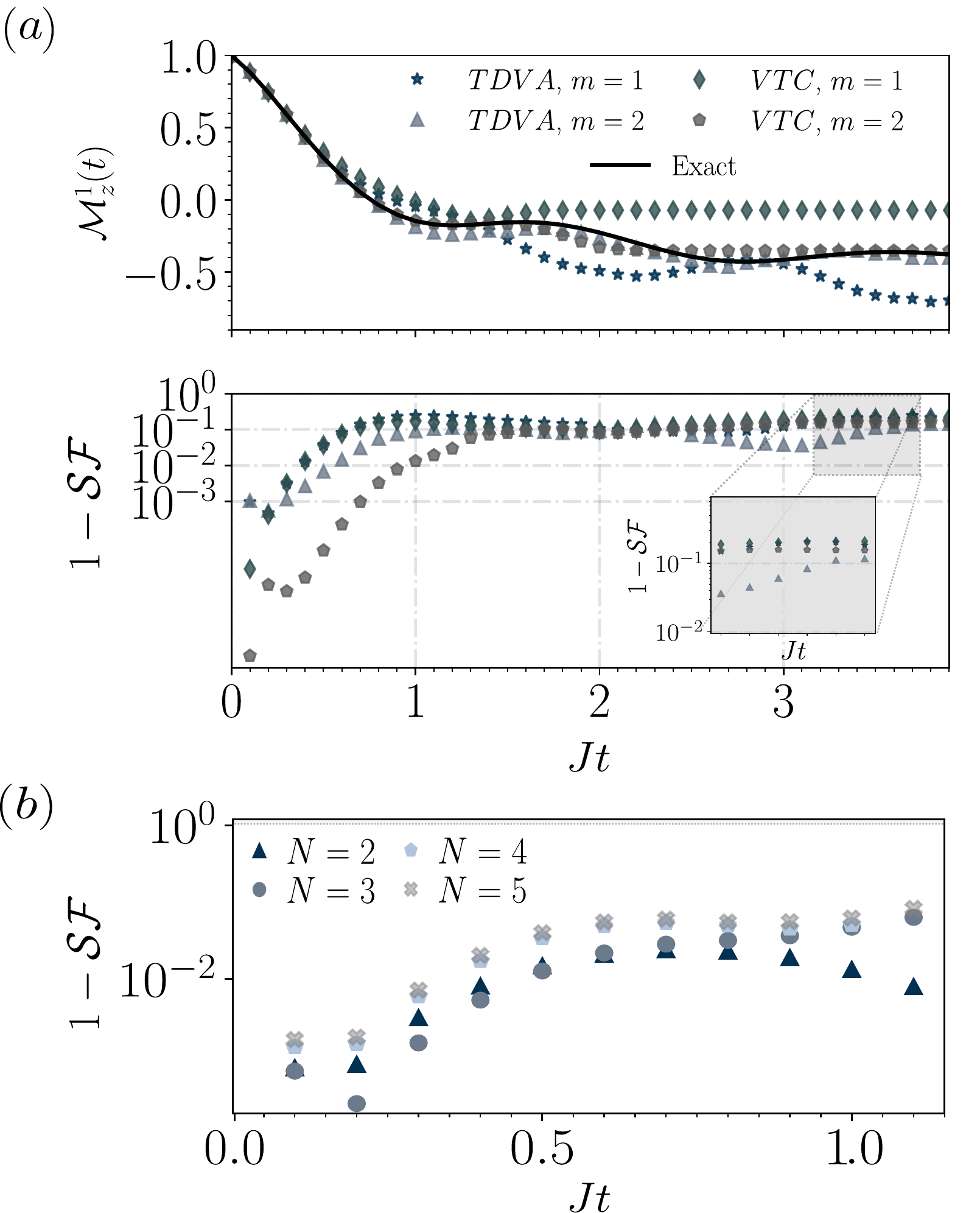}
  \caption{(a) Numerical simulations of variational quantum algorithms for time-evolving a quantum system described by Lindbladian dynamics, whose Hamiltonian is described by Eq.~(\ref{eq:TFI_equation}), and the dissipator, Eq.~(\ref{eq:dissip}), consists of Lindblad operators $\{L_i=\sqrt{\gamma}\sigma_i^-\}_{i=1}^N$ with $\gamma=0.5J$. The initial state $\rho(0)$ is a product state with all spins in the state $|0\rangle$. Top: time evolution of the normalized average magnetization in the $z$ direction at the first site, $\mathcal{M}_z^1$, where $N=3$. We use the variational ansatz $\kettt{\tilde{\rho}^o}$ given in Section \ref{sec:Ansatz_Lindb} with $m=1$ and $2$ and obtained using TDVA and VTC methods. The exact dynamics is plotted as a solid black curve. Bottom: the state infidelity relative to the exact state $\rho$, $1-\mathcal{SF}(\rho,\tilde{\rho}^o)=1-Tr \left[\sqrt{\sqrt{\rho}\tilde{\rho}^o\sqrt{\rho}}\right]^2$, for the different cases studied in the top figure. (b) Performance for different system sizes. We plot the state infidelity for the same setting as in (a), using TDVA method, but with $\gamma=J$. The ansatz chosen here is $\kettt{\tilde{\rho}^o}$ with $m=1$. Tikhonov regularization method is used here to tackle the singularity of $A^R$, Eq.~(\ref{eq:A_C}) (see Appendix \ref{appendix:tikhonov}).}
  \label{fig:open_system_sim}
\end{figure}

In this section, we examine the performance of the algorithm in simulating the dynamics of open quantum systems. We consider Markovian open system dynamics given by Eq.~(\ref{eq:lindblad}) where $H$ is the TFI Hamiltonian given in Eq.~(\ref{eq:TFI_equation}) and $L_i=\sqrt{\gamma}\sigma_i^-=\sqrt{\gamma}(\sigma_i^x-i\sigma_i^y)/2$. with $\gamma=0.5J$. For the simulation shown in Fig.~\ref{fig:open_system_sim}(a), we consider a system composed of three spins that starts from the state $\rho(t=0)=|000\rangle\langle000|$.

\par
In Fig.~\ref{fig:open_system_sim}(a), we show the numerical results obtained using TDVA and VTC methods and employing the ansatz $\kett{\tilde{\rho}^o(\vec{\theta})}$, described in Section \ref{sec:ansatz_construction} with $m=1$ and $m=2$. We additionally require that all of the variational parameters of $G_{i}^g$'s are the same for different $i$'s in the same large Trotter step (see Appendix~\ref{appendix:ansatz construction}). To tackle the singularity issue of $A^R$, arising when implementing the TDVA method, Tikhonov regularization method was implemented. As seen from this figure, using an ansatz with two large Trotter steps gives significantly better results than a single step. 
Fig.~\ref{fig:open_system_sim}(b) shows the system size scaling behavior for this case. Here, the state infidelity is plotted for the time evolution simulation of the same open TFI model described above for different system sizes using TDVA method. In addition to varying the system size, we change the dissipation rate to $\gamma=J$. For all system sizes, the ansatz employed is of the $\kett{\tilde{\rho}^o\big(\vec{\theta}(t)\big)}$ form with one large Trotter step, $m=1$. Interestingly, the state fidelity performance, as shown here, highlights how the ansatz $\ketttt{\tilde{\rho}^o}$ maintains its efficiency in capturing the exact dynamics when considering larger system sizes.
\par
\section{\label{sec:summ}Summary}
In summary, we discussed a variational algorithm for a hybrid quantum-classical computer to simulate both the Lindblad master equation and its adjoint. The algorithm builds upon previously introduced variational quantum algorithms and has similar hardware requirements, up to a linear factor. We further benchmarked the algorithm on two numerical experiments. In the first experiment, we solved the adjoint master equation for time-evolving quantum observables for a closed TFI model reducing this problem to solving the Heisenberg equation of motion. Throughout this experiment, we demonstrated that the ansatz gives an excellent approximation. The second experiment aimed to time-evolve the quantum state using an open version of the TFI model. We showed that a variational state involving non-unitary gates that can be prepared with unitary gates can be used. We also demonstrated that this ansatz gives a good approximation.
\par
\section{\label{sec:Acknowledgements}Acknowledgements}
\textit{Acknowledgements.---} We thank Eyal Bairey for useful discussions. We acknowledge financial support from the Defense Advanced Research Projects Agency through the DRINQS program, grant No. D18AC00025, and from the Israel Science Foundation within the ISF-Quantum program (Grant No. 2074/19). TW acknowledges the support of the Baroness Ariane de Rothschild Women Doctoral Program. 
\appendix

\section{Derivation of the TDVA-Based Dynamics Equation Eq.~(\ref{eq:algebr_eq})}
\label{appendix:deriv_mach}
To derive Eq.~(\ref{eq:algebr_eq}), we start by applying McLachlan's variational principle on the evolution equation Eq.~(\ref{eq:linb_supersp}). McLachlan's principle is obtained by minimizing the distance between the two sides of the evolution equation, i.e., demanding
\begin{equation}
    \delta \norm{\left( \frac{d}{dt}-\mathcal{L} \right)\ketttt{\tilde{\rho}(\vec{\theta})}}=0.
\end{equation}
Or equivalently, $\delta\norm{\left( \frac{d}{dt}+i\mathcal{H}_{SQ} \right)\kettt{\tilde{\rho}(\theta(t))}}^2=0$, where $\mathcal{H}_{SQ}=i\mathcal{L}$ is what we defined to be the super Hamiltonian. From here, we obtain,
\begin{equation}
    \delta\left[ \left( \left( \frac{d}{dt}+i\mathcal{H}_{SQ}\right)\ketttt{\tilde{\rho}}\right)^\dagger \left( \left( \frac{d}{dt}+i\mathcal{H}_{SQ}\right)\ketttt{\tilde{\rho}}\right) \right]=0 
\end{equation}
Substituting $\frac{d}{dt}\kett{\tilde{\rho}\big(\vec{\theta}(t)\big)} =\sum_j{\dot{\theta}_j\frac{d}{d\theta_j}\kett{\tilde{\rho}\big(\vec{\theta}(t)\big)}}$ yields
\begin{equation}
\begin{aligned}
    &\delta\left[ \left( \sum_j {\dot{\theta}_j \frac{d\bbbbra{\tilde{\rho}}}{d\dot{\theta}_j}}-i\bbbbra{\tilde{\rho}}{\mathcal{H}_{SQ}}^\dagger \right)\left(\sum_i {\dot{\theta}_i \frac{d\ketttt{\tilde{\rho}}}{d\dot{\theta}_i}}+i\mathcal{H}_{SQ}\ketttt{\tilde{\rho}} \right) \right]\\
    &=0 
\end{aligned}
\end{equation}
Rearranging this equation gives
\begin{equation}    
\begin{aligned}
&\delta \bigg[\sum_{ij}{\dot{\theta}_j \dot{\theta}_i\overlaps{\partial_j\tilde{\rho}}{\partial_i\tilde{\rho}}}+i\sum_j{\dot{\theta}_j \bbbbra{\partial_j\tilde{\rho}}\mathcal{H}_{SQ}\ketttt{\tilde{\rho}}}\\
&-i\sum_i{\dot{\theta}_i \bbbbra{\tilde{\rho}}{\mathcal{H}_{SQ}}^\dagger\ketttt{\partial_i\tilde{\rho}}} 
+\bbbbra{\tilde{\rho}}{\mathcal{H}_{SQ}}^\dagger\mathcal{H}_{SQ}\ketttt{\tilde{\rho}} \bigg] =0
\end{aligned}
\end{equation}
We finally have 
\begin{equation}
\begin{aligned}
    \delta \|\cdot\|^2=&\sum_i\delta\theta_i\bigg[\sum_j {\dot{\theta}_j\left( \overlaps{\partial_i\tilde{\rho}}{\partial_j\tilde{\rho}}+\overlaps{\partial_j\tilde{\rho}}{\partial_i\tilde{\rho}}\right)}+\\
    &i\left(\bbbbra{\partial_i\tilde{\rho}}\mathcal{H}_{SQ}\ketttt{\tilde{\rho}}-\bbbbra{\tilde{\rho}}{\mathcal{H}_{SQ}}^\dagger\ketttt{\partial_i\tilde{\rho}}\right)\bigg]=0 \iff\\
    &\sum_j {Re \ \overlaps{\partial_i\tilde{\rho}}{\partial_j\tilde{\rho}}\dot{\theta_j}}=Im \bbbbra{\partial_i\tilde{\rho}}\mathcal{H}_{SQ}\ketttt{\tilde{\rho}} 
\end{aligned}
\end{equation}

This equation holds for $\mathcal{H}_{SQ}$, which is non-Hermitian. For the dynamics problem of quantum observables governed by the evolution equation Eq.~(\ref{eq:adj_mstr}), we have
\begin{equation}
        \sum_j {Re \ \overlaps{\partial_i\tilde{O}}{\partial_j\tilde{O}}\dot{\theta_j}}=Im \bbbbra{\partial_i\tilde{O}}\mathcal{H}_{SQ}^\dagger\ketttt{\tilde{O}} 
\end{equation}
with $\mathcal{H}_{SQ}^{\dagger}=i\mathcal{L}^{\dagger}$.
In addition to McLachlan's principle, the proposed algorithm can be based on other quantum time evolution principles, such as the time-dependent variational principle (TDVP) as demonstrated in \cite{yuan2019theory}. 
The algebraic evolution equations obtained by TDVP are 
\begin{equation}
    \sum_j {Im \ \overlaps{\partial_i\tilde{\rho}}{\partial_j\tilde{\rho}}\dot{\theta_j}}=-Re \bbbbra{\partial_i\tilde{\rho}}\mathcal{H}_{SQ}\ketttt{\tilde{\rho}}.
\end{equation}
Throughout this paper, we choose McLachlan's principle over TDVP since for closed systems, at the starting point of the time evolution, i.e., $t=0$, the elements $\bbra{\partial_i\tilde{O}}\mathcal{H}_{SQ}\kett{\tilde{O}}$ for Trotter-like ansatze defined in Eq.~(\ref{eq:Trotter-like ansatz}) are pure imaginary. Therefore, the TDVP-based algorithm fails since the right side of the equation is zero. 

\section{Ansatz Construction - Technical Details}
\label{appendix:ansatz construction}
As an example of constructing ansatze of the Trotter-like form discussed in the paper, here we demonstrate and detail the procedure for constructing the $\kett{\tilde{O}_1^c(\theta)}$ and $\kettt{\tilde{\rho}^o(\theta)}$ ansatze used in the first and second numerical experiments, respectively, with $m=2$. 
\par
\begin{figure}[]
\includegraphics[width=\linewidth, trim={0cm 2.5cm 5cm 0cm},clip]{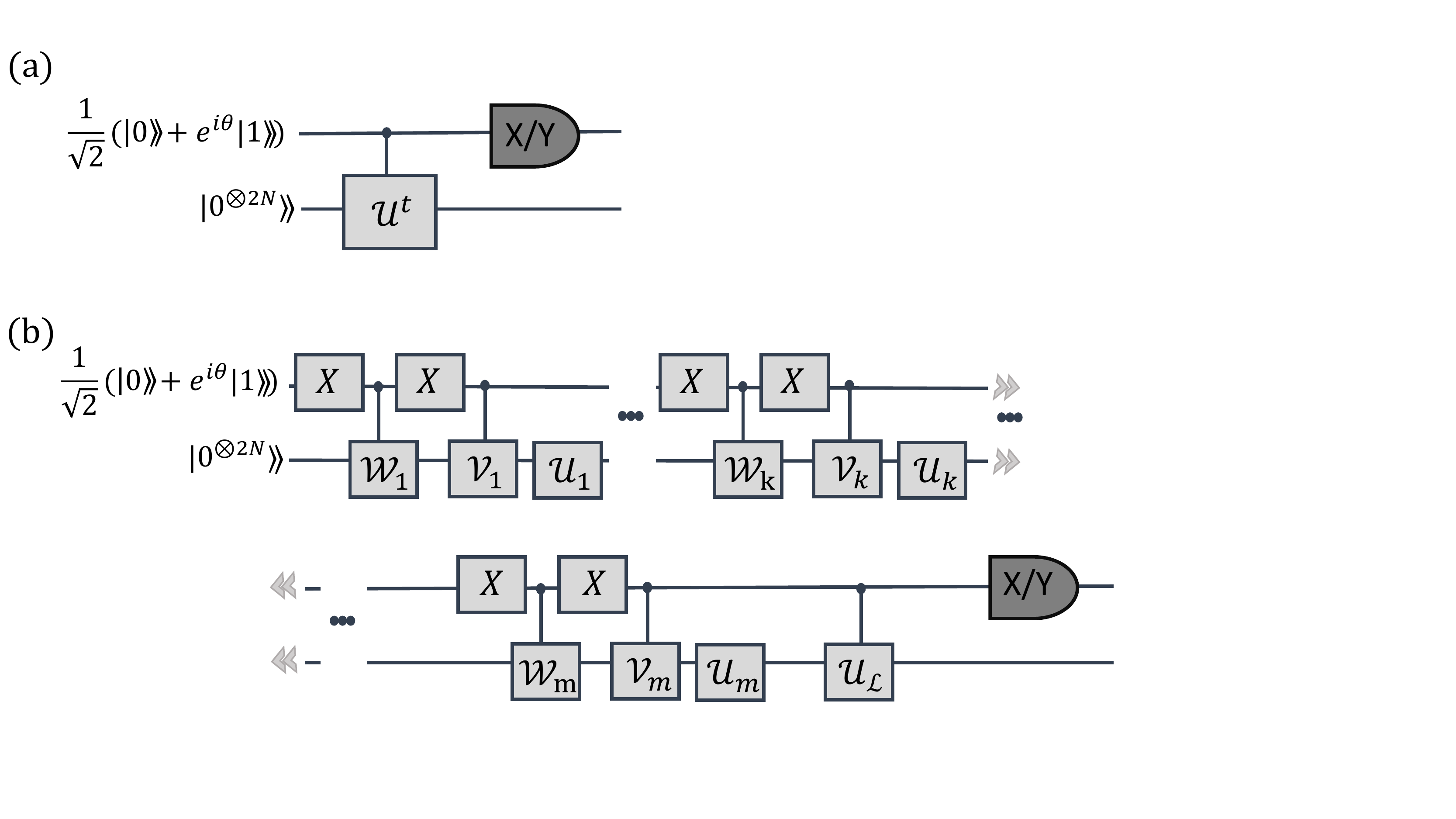}
  \caption{(a) Quantum Circuit implementing the Hadamard test that can be used for evaluating the expectation value $\bbbra{0^{\otimes 2N }}\mathcal{U}^t\kettt{0^{\otimes 2N}}$, where $\mathcal{U}^t$ is unitary. (b) A simplified version of (a).}
\label{fig:hadamard_test_simplified_version}
\end{figure}

The model used in the first numerical experiment is the closed $TFI$ model, Eq.~(\ref{eq:TFI_equation}). For simplicity, let us consider the $N=2$ case, where the physical system is composed of $2$ spins. As explained in the introduction section, the first step of the algorithm is to map the problem to that of time-evolving a larger qubit system state $\kettt{O(0)}$ consisting of $4$ qubits. This evolution is governed by the super Hamiltonian $\mathcal{H}^\dagger_{SQ}$, which is represented in the standard basis of the new super qubit system and is given by $\mathcal{H}^\dagger_{SQ}=i\mathcal{L}_H^\dagger$. In the new basis, $\mathcal{L}_H^\dagger$ can be written as follows
\begin{figure*}[]
\includegraphics[width=0.8\linewidth, trim={0cm 4cm 1cm 1cm},clip]{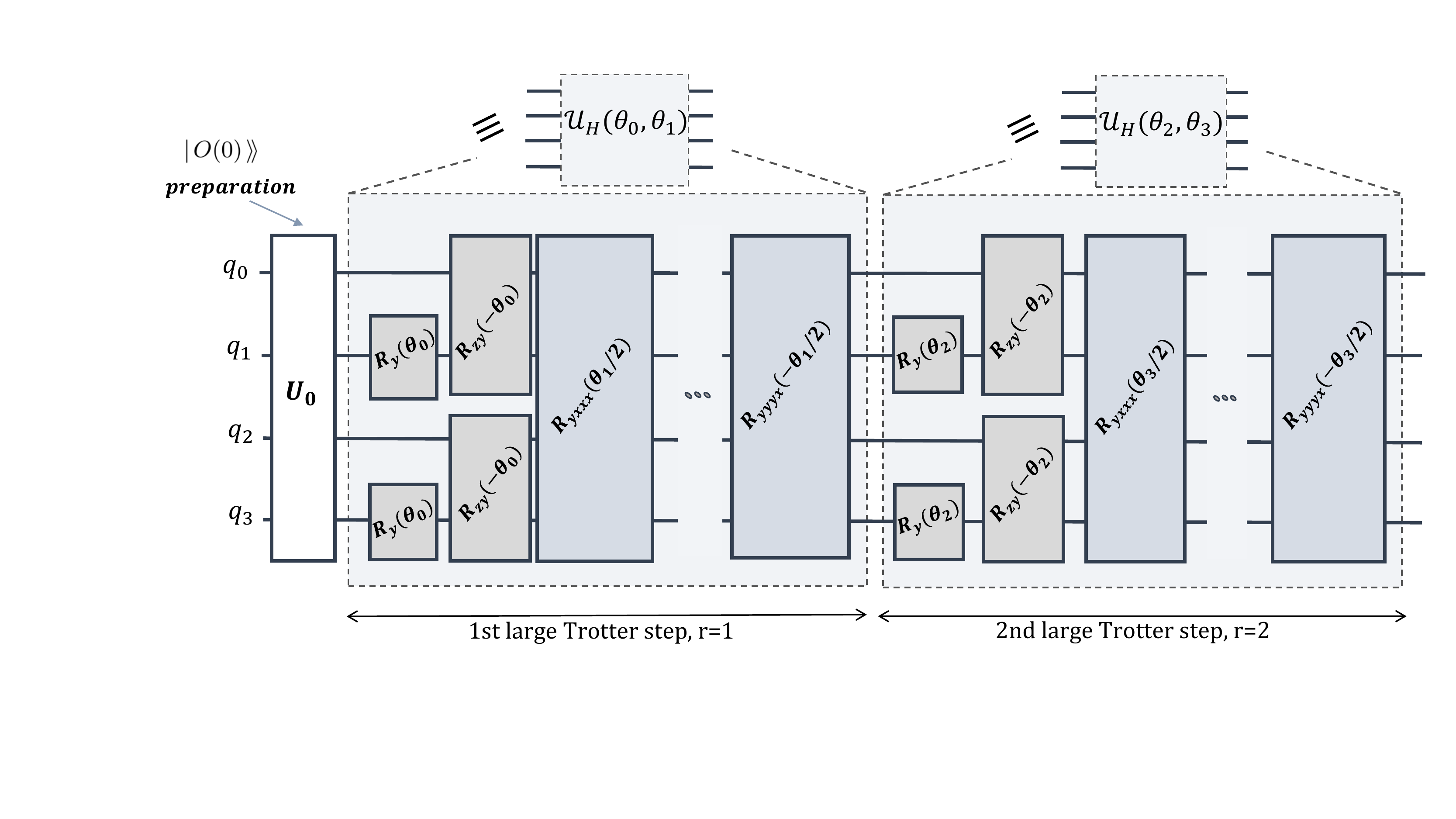}
  \caption{Quantum Circuit for preparing ansatz $\kett{\tilde{O}_1^c \bigl(\vec{\theta}(t)\bigl)}$ for a system composed of 2 spins described by the closed $TFI$ model. The ansatz illustrated here is composed of two large Trotter steps, i.e., $m=2$.}. 
  \label{fig:ansatz_o1_struct}
\end{figure*}
\begin{figure*}[]
\includegraphics[width=0.8\linewidth, trim={0cm 0cm 1cm 0cm},clip]{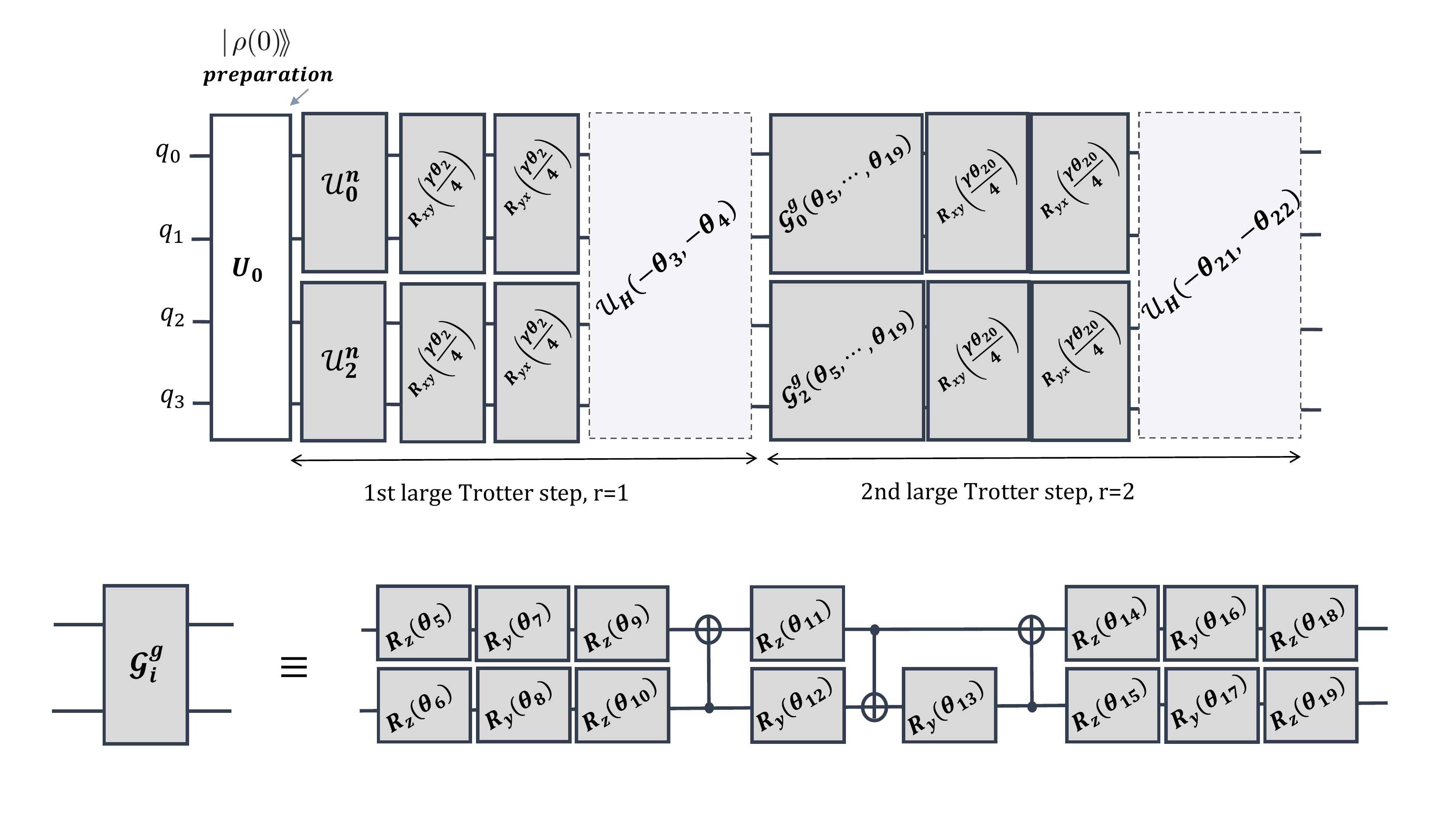}
  \caption{Quantum Circuit for preparing ansatz $\kett{\tilde{\rho}^o\bigl(\vec{\theta}(t)\bigl)}$ used in the second numerical experiment for a system composed of 2 spins and is described by the open $TFI$ model. The ansatz is composed of two large Trotter steps, i.e., $m=2$. The circuit $\mathcal{U}_H$ is defined in Fig.~\ref{fig:ansatz_o1_struct}, and a possible decomposition of a generic two-qubit gate $\in U(2)$, $\mathcal{G}_i^g$, into single-qubit gates and CNOT gates is shown here. This scheme is based on the approach introduced in \cite{PhysRevA.69.032315}.}
  \label{fig:ansatz_o1_struct_open}
\end{figure*} 
\begin{equation}
\label{eq:ex_liouvillian}  \mathcal{L}_H^\dagger=ih\sum_{i=0}^1{\left(\sigma^y_{2i+1}-\sigma^z_{2i}\sigma^y_{2i+1}\right)}+i\frac{J}{2}\sum_{\{\vec{\tau}\}}{c_{\tau} \sigma^{\tau^0}_0\sigma^{\tau^1}_1\sigma^{\tau^2}_2\sigma^{\tau^3}_3},
\end{equation}
where $\{\vec{\tau}=(\tau^0, \tau^1, \tau^2, \tau^3)\}$ includes all the permutations of $(y,x,x,x)$ and $(x,y,y,y)$, and $c_{\tau}$ is the parity of the permutations. The first sum corresponds to the external field part, and the second corresponds to the $ZZ$ interaction term of the original Hamiltonian.   
Recall that $\kett{\tilde{O}_1^c(\theta)}$, given the decomposition of $\mathcal{L}$ in Eq.~(\ref{eq:op_liouv}), is defined as 
\begin{equation}
\label{eq:ansatz_closed_appendix}
    \kett{\tilde{O}_1^c\big(\vec{\theta}(t)\big)}= \Pi_{r=1}^m \Pi_{i:even,\vec{\alpha}} {\mathcal{U}_{i,\vec{\alpha}}^h \left( \theta_{i,r}^{\vec{\alpha}}\right)}\kettt{O(0)},
\end{equation}
where $\mathcal{U}_{i,\vec{\alpha}}^h\left( \theta_{i,r}^{\vec{\alpha}}\right) =\exp\left(\theta_{i,r}^{\vec{\alpha}} (c^h_{i,{\vec{\alpha}}} \Lambda^h_{i,{\vec{\alpha}}})^\dagger\right)$, and all the gates within the same large Trotter steps are divided into commutative subgroups. Additionally, a similar variational parameter is assigned for each subgroup gates elements.  Explicitly, $c^h_{i,{\vec{\alpha}}}$ is non trivial for $\vec{\alpha}=(0,y,0,0)$, $\vec{\alpha}=(0,0,0,y)$, $\vec{\alpha}=(z,y,0,0)$, $\vec{\alpha}=(0,0,z,y)$ and $\vec{\alpha}\in \{\vec{\tau}\}$. Moreover, all the Pauli terms in $\mathcal{L}_H^\dagger$ corresponding to the $ZZ$ interaction term of the original $TFI$ Hamiltonian commute with each other. As a result, the same variational parameter is used for all these terms within each large Trotter step. This property also holds for the Pauli terms corresponding to the external field term. Given this, the total number of different parameters will be $2m$. Putting all together, $\kett{\tilde{O}_1^c(\theta)}$ with $m=2$ simplifies to
\begin{equation}
\label{ansatz_closed_rot}
\begin{aligned}
        \kett{\tilde{O}_1^c\big(\vec{\theta}(t)\big)}= &\biggl(\Pi_{\vec{\tau}}{R_{\vec{\tau}}^{0,1,2,3}\left(\frac{\theta_3}{2}\right)}R_{(z,y)}^{2,3}(-\theta_2)R_{(z,y)}^{0,1}(-\theta_2)\\&R_y^3(\theta_2)
        R_y^1(\theta_2)\biggl)\cdot\biggl(\Pi_{\vec{\tau}}{R_{\vec{\tau}}^{0,1,2,3}\left(\frac{\theta_1}{2}\right)}R_{(z,y)}^{2,3}(-\theta_0)\\&R_{(z,y)}^{0,1}(-\theta_0)R_y^3(\theta_0) R_y^1(\theta_0)\biggl)\kettt{O(0)}\\&\equiv \mathcal{U}_{H}(\theta_2,\theta_3)\mathcal{U}_{H}(\theta_1,\theta_0)\kettt{O(0)},
\end{aligned}
\end{equation}
where $R_{(\alpha_{i_0}, \alpha_{i_1},\cdots)}^{i_0,i_1,\cdots}(\theta)=\exp\left(i\theta\sigma_{i_0}^{\alpha_{i_0}}\sigma_{i_1}^{\alpha_{i_1}}\cdots\right)$ and $i_0,i_1,\cdots \in \{0,1,2,3\}$.
The quantum circuit for preparing this ansatz is illustrated in Fig.~\ref{fig:ansatz_o1_struct}(a).
Considering this ansatz, $A^R$ and $C^I$ matrices elements, can be rewritten as a combination of expectation values of the form 
\begin{equation}
\label{eq:expect_ut}
\begin{aligned}
    \bbbra{0^{\otimes 2N }}\mathcal{U}^t\kettt{0^{\otimes 2N}}=& \bbbra{0^{\otimes 2N }}\left( \otimes_{k=1}^m \mathcal{W}^\dagger_{m-k+1}\mathcal{U}_{m-k+1}^\dagger\right)\cdot\\&\mathcal{U}_{\mathcal{L}}\cdot\left(\otimes_{k=1}^m \mathcal{U}_{k}\mathcal{V}_k\right)\kettt{0^{\otimes 2N}},
\end{aligned}
\end{equation}
Where in order to improve readability, we refer to the set of gates $\mathcal{U}_{i,\vec{\alpha}}^h\left(\theta_{i,r}^{\vec{\alpha}}\right)$'s in Eq.~(\ref{eq:ansatz_closed_appendix}) with a single index $k$ as $\mathcal{U}_k$'s. The unitaries $\mathcal{V}_k$ and $\mathcal{W}^\dagger_k$ are both either trivial or equal to the derivative of the rotation gate $\mathcal{U}_k$ and its conjugate transpose, respectively, up to a complex factor. Specifically, for the $C^I$ elements only one of the $\mathcal{W}^\dagger_k$'s will be non-trivial and for the $A^R$ elements, only one of the $\mathcal{W}^\dagger_k$'s and one of $\mathcal{V}_k$'s will be non-trivial. $\mathcal{U}_{\mathcal{L}}$ is either trivial for the terms corresponding to $A^R$ elements or one of the set $\{\Lambda_{i\vec{\alpha}}^h, \Delta_{i\vec{\beta}}^u, \Delta_{i\vec{\gamma}}^n\}$'s for terms corresponding to $C^I$ elements.
The expectation value of $\mathcal{U}^t$,  can be evaluated using a Hadamard test. The circuit implementing the Hadamard test is shown in Fig. \ref{fig:hadamard_test_simplified_version}(a). We can determine the expectation value of $\mathcal{U}_t$ by measuring the probability of finding the ancillary qubit in the $\kett{+x}$ and $\kett{+y}$ states. These probabilities are denoted by $p(\kett{+y})$ and $p(\kett{+x})$, respectively. Using the following relations,

\begin{equation}
    \label{eq:had_test_relations}
    \begin{aligned}        
    &Re\left(e^{i\theta} \bbbra{0^{\otimes 2N }}\mathcal{U}^t\kettt{0^{\otimes 2N}}\right)=2p(\kett{+x})-1\\& Im\left(e^{i\theta} \bbbra{0^{\otimes 2N }}\mathcal{U}^t\kettt{0^{\otimes 2N}}\right)=2p(\kett{+y})-1.
    \end{aligned}
\end{equation}

In addition, it is possible to evaluate such terms using a simplified quantum circuit with a shorter depth. This circuit requires considerably fewer controlled gates than the standard Hadamard test circuit. Such circuit is presented and explained in detail in \cite{li2017efficient, ekert2002direct} and is shown in Fig. \ref{fig:hadamard_test_simplified_version}(b). The output state of the system and the ancilla qubit of this circuit is
\begin{equation}
    \begin{aligned}
    \frac{1}{\sqrt{2}}&\bigl(\ketttt{0}\mathcal{U}_m\mathcal{W}_m\cdots \mathcal{U}_2\mathcal{W}_2\mathcal{U}_1\mathcal{W}_1\kettt{0^{\otimes 2N}}+\\&e^{i\theta}\ketttt{1}\mathcal{U}_m\mathcal{V}_m\cdots \mathcal{U}_2\mathcal{V}_2\mathcal{U}_1\mathcal{V}_1\kettt{0^{\otimes 2N}}\bigr).
    \end{aligned}
\end{equation}
Which can be rewritten as 
\begin{equation}
    \begin{aligned}
    \frac{1}{\sqrt{2}}&\mathcal{U}_m\mathcal{W}_m\cdots \mathcal{U}_2\mathcal{W}_2\mathcal{U}_1\mathcal{W}_1\bigl(\ketttt{0}\kettt{0^{\otimes 2N}}+\\&e^{i\theta}\ketttt{1}\mathcal{W}_1^\dagger \mathcal{U}_1^\dagger\cdots \mathcal{W}_m^\dagger \mathcal{U}_m^\dagger \mathcal{U}_m\mathcal{V}_m\cdots \mathcal{U}_1\mathcal{V}_1\kettt{0^{\otimes 2N}}\bigl)
    \\&=\frac{1}{\sqrt{2}}\mathcal{U}_m\mathcal{W}_m\cdots \mathcal{U}_1\mathcal{W}_1\bigl(\ketttt{0}\kettt{0^{\otimes 2N}}+e^{i\theta}\ketttt{1}\mathcal{U}_t \kettt{0^{\otimes 2N}}\bigl).
    \end{aligned}
\end{equation}
Therfore, by measuring the ancillary qubit in basis $X$ and $Y$, we can evaluate the expectation values given in Eq.~(\ref{eq:expect_ut}) using the same relations given in Eq.~(\ref{eq:had_test_relations}).
For simulations based on VTC method, it is possible to measure the overlap involved in the cost function, Eq.~(\ref{eq:VTC_fidelity}), directly on a quantum computer by assuming $\kett{\tilde{\rho}(\vec{\theta}(t))}=A(\vec{\theta}(t))\kettt{0^{\otimes 2N }}$, preparing the state $A^{\dagger}(\vec{\theta}(t+\delta t))\hat{V}(\delta t)A(\vec{\theta}(t))\kettt{0^{\otimes 2N }}$ and measuring the probability that all the qubits end up in the state, $\kettt{0^{\otimes 2N }}$.
Alternatively, a swap-test can be implemented, which requires carrying a measurement only on the ancillary qubit but uses $2N+1$ qubits \cite{buhrman2001quantum, foulds2021controlled}. Another option is the Hadamard test which requires less qubits and measurement of one ancillary qubit but is based on a deeper circuit than the former. Also, unlike the first option, for the implementation of the SWAP-test and Hadamard test, controlled gates are required.
\par Similarly, the Liouvillian in the new basis corresponding to the open version of the $TFI$ model, with the single site jump operators, $L_i=\sqrt{\gamma}\sigma_i^{-}$'s, considered in the second numerical experiment, is 
\begin{equation}
\label{eq:ex_liouvillian_open}
    \begin{aligned}
        \mathcal{L}=&-\mathcal{L}_H^\dagger-\sigma_0^0\sigma_1^0\sigma_2^0\sigma_3^0+\frac{\gamma}{4}\sum_{i=0}^3 \sigma^z_{i}-\\&\frac{\gamma}{4}\sum_{i=0}^1 \biggl(\sigma^x_{2i}\sigma^x_{2i+1}-\sigma^y_{2i}\sigma^y_{2i+1} -i\sigma^x_{2i}\sigma^y_{2i+1}-i\sigma^y_{2i}\sigma^x_{2i+1}\biggl),
    \end{aligned}
\end{equation}
where $\mathcal{L}_H^\dagger$ is given in Eq.~(\ref{eq:ex_liouvillian}).
\par The quantum ansatz, considered in the second numerical experiment and defined in Section \ref{sec:ansatz_construction}, is of the form 
\begin{equation}
\label{eq:open_system_trotter_ansatz}
\begin{aligned}
&\ketttt{\tilde{\rho}^o(\vec{\theta})}=\Pi_{r=1}^m \biggl[ \biggl(\Pi_{i:even}\Pi_{\vec{\alpha}}{\mathcal{U}_{i,\vec{\alpha}}^h \big( \theta_{i,r}^{\vec{\alpha}}\big)}\biggl)\cdot\\ &\biggl(\Pi_{i:even}\Pi_{\vec{\beta}}{\mathcal{U}_{i,{\vec{\beta}}}^u \big( \theta_{i,r}^{\vec{\beta}}\big)}\biggl)\cdot \Pi_{i:even} \mathcal{F}_{i,r}\biggr]\kettt{\rho(0)}
\end{aligned}
\end{equation}
where the variational gate $\mathcal{F}_{i,r}$, is given by 
\begin{enumerate}[(i)]
    \item For $r=1$, $\mathcal{F}_{i,1}=\Pi_{\vec{\gamma}}\mathcal{G}_{i,\vec{\gamma}}^n(\theta_{i,1}^{\vec{\gamma}}).$ As this is a non-unitary gate, this gate's action on $\kettt{\rho(0)}$ on a quantum computer is simulated using a unitary gate $\mathcal{U}^n_i$ that satisfies $\mathcal{F}_{i,1}\kettt{\rho(0)} = a_{i, 1}\cdot\mathcal{U}^n_i\kettt{\rho(0)}$, and $a_{i,1}$ is a complex factor. 
    \item For subsequent layers, $r>1$, $\mathcal{F}_{i,r} = a_{i,r}\cdot\mathcal{G}^g_i$ where $\mathcal{G}^g_i$ represents variational two-qubit gates applied to even pairs of qubits. We use a parameterization in which all $\mathcal{G}^g_i$'s gates have a similar structure. In the second experiment, we further require that all the $G_i^g$'s gates have the same set of variational parameters $\vec{\theta}_r^g$ for all $i$. The complex factors $a_{i,r}$ vary for each qubit $i$. 
\end{enumerate}
An illustration for this ansatz for the open $TFI$ model is shown in Fig.~\ref{fig:ansatz_o1_struct_open} with $m=2$.
To achieve this form, the sequence of non-unitary gates acting on the first pair of qubits in the first large Trotter layer of the ansatz, 
\begin{equation}
    \begin{aligned}
    &\mathcal{G}^n_{0,\vec{\gamma}=(x,x,0,0)}\mathcal{G}^n_{0,\vec{\gamma}=(y,y,0,0)}\mathcal{G}^n_{0,\vec{\gamma}=(z,0,0,0)}\mathcal{G}^n_{0,\vec{\gamma}=(0,z,0,0)}=\\&R^{0,1}_{(x,x)}(i\gamma /4\theta_1)\cdot R^{0,1}_{(y,y)}\left(-i\gamma /4\theta_1\right)\cdot R^{0}_{z}\left(-i\gamma /4\theta_0\right)\\&\cdot R^{1}_{z}\left(-i\gamma /4\theta_0\right),
    \end{aligned}
\end{equation}
together with the identity dissipation term, $\exp\left(-t\sigma_0^0\sigma_1^0\sigma_2^0\sigma_3^0\right)$, is simulated using a unitary circuit, $\mathcal{U}_0^n$. Note that commutative gates have similar variational parameters. Similarly, we find $\mathcal{U}_2^n$ which simulates the action of all the non-unitary gates acting on the second pair of qubits. $\mathcal{U}_2^n$ can be different from $\mathcal{U}_0^n$ as the initial states of the two pairs of qubits may differ. Moreover, we find a set of unitaries simulating the derivatives of this sequence of non-unitary gates with respect to all the different variational phases. Handling the unitary gates is similar to the closed system case. In the second large Trotter layer, every sequence of non-unitaries acting on a pair of qubits is replaced by a general unitary up to a complex factor. For the simulation shown in Section \ref{sec:NE-open}, we assumed that general unitaries share the same set of parameters.  The obtained ansatz and a possible choice for the general unitary are shown in Fig.~\ref{fig:ansatz_o1_struct_open}. To evaluate the elements $A^R$ and $C^I$, similar circuits to those discussed for the closed system case can be implemented, including the simplified version with slight modifications.

\section{Tikhonov Method}
\label{appendix:tikhonov}
A main serious concern arises when matrix $A^R$ in Eq.~(\ref{eq:algebr_eq}) is ill-conditioned. In particular, for ansatze with complex structures, it can be strictly singular, and therefore, it is necessary to employ regularization methods to find a stable solution for this equation. One of the most widely used variational regularization methods for obtaining approximate solutions for such problems is the Tikhonov regularization method. Tikhonov regularization was implemented in the simulations shown in Fig.~ \ref{fig:First_and_Second_Figures_Combined3} and Fig. \ref{fig:open_system_sim} where the main idea is as follows. Let $A$ be an $m\times n$ matrix with rank $r$ an $b\in \mathbb{R}^m$ a given vector. The least squares solution of the linear equations $Ax=b$ is a minimization problem that aims to find an approximate solution $x_{LSM}$ that satisfies
\begin{equation}
    x_{LSM}=\min_{x\in \mathbb{R}^n} \norm{Ax-b}_2^2
\end{equation}
where $\norm{\cdot}_2$ denotes the 2-norm (Euclidean norm). Solving this problem by substituting the SVD decomposition of the $A$ matrix, $A=\sum_{i=1}^r \sigma_i u_i v_i^T$ where $u_1,u_2,\cdots u_m$ and $v_1,v_2,\cdots v_n$ are two sets of orthonormal bases and $\sigma_i$'s are the singular values, yields
\begin{equation}
    x_{LSM}=\sum_{i=1}^r {\frac{u_i^T b v_i}{\sigma_i}}.
\end{equation}
Such a solution is contaminated by the noise in directions corresponding to small singular values of the matrix $A$ and, therefore, can be very unstable, and its norm can be very large. A small change in vector $b$ can lead to a large change in the solution vector. With the hope to filter the effect of the noise out, Tikhonov proposed to replace this problem with the penalized minimization problem 
\begin{equation}
    x_{TIK}=\min_{x\in \mathbb{R}^n} \norm{Ax-b}_2^2+\lambda^2 \norm{x}_2^2
\end{equation}
Again, using the SVD decomposition of the matrix $A$, the solution to the modified minimization problem is 
\begin{equation}
    x_{TIK}=\sum_{i=1}^r {\frac{\sigma_i}{\sigma_i^2+\lambda^2} u_i^T b v_i}.
\end{equation}
Notice that the obtained solution is similar to $x_{LSM}$, but each term in the summation is multiplied by a filter factor which is equal to $\frac{\sigma_i^2}{\sigma_i^2+\lambda^2}$. The solution efficiency depends on the choice of the regularization parameter, $\lambda$, that controls a trade-off between minimizing the 2-norm of $x$, and hence the solution being stable, and how small the residual $\norm{Ax-b}_2$ is which defines how accurate the solution is. An efficient way to determine the regularization parameter is using what is known as the L-curve  \cite{lawson1974solving, hansen1993use, hansen1992analysis}. L-curve is a plot of the points $\left( \rho(\lambda), \eta(\lambda)\right)$ where $\rho(\lambda)=\norm{Ax_{TIK}-b}_2$ and $\eta(\lambda)=\norm{x_{TIK}}_2$. The L-curve has a shape similar to the letter L where the chosen regularization parameter is chosen to be the one corresponding to the corner of the $L-$shaped curve. A justification for this is that the flat part of this shape corresponds to when the solution $x_{TIK}$ is affected by regularization errors, and the vertical part of the L-curve is affected by errors due to noise etc. Thus, a good choice that gives a good balance between the accuracy and the stability of the solution is the corner. However, sometimes the curve does not have one corner, and hence it is recommended to choose the corner which is the leftmost of some bound, $M$, on the $2-norm$ of the solution.
\par
Another alternative method that can be implemented is the truncated SVD (TSVD) method, according to which the solution is given by $x_{TSVD}=A^{-1}_k b$ where $A^{-1}_k=V\Sigma_k^{-1}U^T$. $A^{-1}_k$ is the pseudoinverse of $A_k=\sum_{i=1}^k {\sigma_i u_i v_i^T}$  and $\Sigma_k^{-1}\in R^{n\times m}$ is a diagonal matrix with the values $\{ 1/\sigma_1, 1\sigma_2, \cdots, 1/\sigma_k, 0 \cdots, 0, 0 \}$ on the diagonal.
Moreover, as in the Tikhonov method, regularization methods can be implemented to choose where the SVD decomposition should be truncated, i.e., $k$.
\bibliography{References} 

\begin{thebibliography}{38}%
\makeatletter
\providecommand \@ifxundefined [1]{%
 \@ifx{#1\undefined}
}%
\providecommand \@ifnum [1]{%
 \ifnum #1\expandafter \@firstoftwo
 \else \expandafter \@secondoftwo
 \fi
}%
\providecommand \@ifx [1]{%
 \ifx #1\expandafter \@firstoftwo
 \else \expandafter \@secondoftwo
 \fi
}%
\providecommand \natexlab [1]{#1}%
\providecommand \enquote  [1]{``#1''}%
\providecommand \bibnamefont  [1]{#1}%
\providecommand \bibfnamefont [1]{#1}%
\providecommand \citenamefont [1]{#1}%
\providecommand \href@noop [0]{\@secondoftwo}%
\providecommand \href [0]{\begingroup \@sanitize@url \@href}%
\providecommand \@href[1]{\@@startlink{#1}\@@href}%
\providecommand \@@href[1]{\endgroup#1\@@endlink}%
\providecommand \@sanitize@url [0]{\catcode `\\12\catcode `\$12\catcode
  `\&12\catcode `\#12\catcode `\^12\catcode `\_12\catcode `\%12\relax}%
\providecommand \@@startlink[1]{}%
\providecommand \@@endlink[0]{}%
\providecommand \url  [0]{\begingroup\@sanitize@url \@url }%
\providecommand \@url [1]{\endgroup\@href {#1}{\urlprefix }}%
\providecommand \urlprefix  [0]{URL }%
\providecommand \Eprint [0]{\href }%
\providecommand \doibase [0]{http://dx.doi.org/}%
\providecommand \selectlanguage [0]{\@gobble}%
\providecommand \bibinfo  [0]{\@secondoftwo}%
\providecommand \bibfield  [0]{\@secondoftwo}%
\providecommand \translation [1]{[#1]}%
\providecommand \BibitemOpen [0]{}%
\providecommand \bibitemStop [0]{}%
\providecommand \bibitemNoStop [0]{.\EOS\space}%
\providecommand \EOS [0]{\spacefactor3000\relax}%
\providecommand \BibitemShut  [1]{\csname bibitem#1\endcsname}%
\let\auto@bib@innerbib\@empty
\bibitem [{\citenamefont {Preskill}(2018)}]{preskill2018quantum}%
  \BibitemOpen
  \bibfield  {author} {\bibinfo {author} {\bibfnamefont {J.}~\bibnamefont
  {Preskill}},\ }\href {\doibase https://doi.org/10.22331/q-2018-08-06-79}
  {\bibfield  {journal} {\bibinfo  {journal} {Quantum}\ }\textbf {\bibinfo
  {volume} {2}},\ \bibinfo {pages} {79} (\bibinfo {year} {2018})}\BibitemShut
  {NoStop}%
\bibitem [{\citenamefont {Feynman}(1982)}]{Feynman1982}%
  \BibitemOpen
  \bibfield  {author} {\bibinfo {author} {\bibfnamefont {R.~P.}\ \bibnamefont
  {Feynman}},\ }\href {\doibase https://doi.org/10.1007/BF02650179} {\bibfield
  {journal} {\bibinfo  {journal} {International Journal of Theoretical
  Physics}\ }\textbf {\bibinfo {volume} {21}},\ \bibinfo {pages} {467–488}
  (\bibinfo {year} {1982})}\BibitemShut {NoStop}%
\bibitem [{\citenamefont {Lloyd}(1996)}]{10.2307/2899535}%
  \BibitemOpen
  \bibfield  {author} {\bibinfo {author} {\bibfnamefont {S.}~\bibnamefont
  {Lloyd}},\ }\href {http://www.jstor.org/stable/2899535} {\bibfield  {journal}
  {\bibinfo  {journal} {Science}\ }\textbf {\bibinfo {volume} {273}},\ \bibinfo
  {pages} {1073} (\bibinfo {year} {1996})}\BibitemShut {NoStop}%
\bibitem [{\citenamefont {Berry}\ \emph {et~al.}(2015)\citenamefont {Berry},
  \citenamefont {Childs}, \citenamefont {Cleve}, \citenamefont {Kothari},\ and\
  \citenamefont {Somma}}]{PhysRevLett.114.090502}%
  \BibitemOpen
  \bibfield  {author} {\bibinfo {author} {\bibfnamefont {D.~W.}\ \bibnamefont
  {Berry}}, \bibinfo {author} {\bibfnamefont {A.~M.}\ \bibnamefont {Childs}},
  \bibinfo {author} {\bibfnamefont {R.}~\bibnamefont {Cleve}}, \bibinfo
  {author} {\bibfnamefont {R.}~\bibnamefont {Kothari}}, \ and\ \bibinfo
  {author} {\bibfnamefont {R.~D.}\ \bibnamefont {Somma}},\ }\href {\doibase
  10.1103/PhysRevLett.114.090502} {\bibfield  {journal} {\bibinfo  {journal}
  {Phys. Rev. Lett.}\ }\textbf {\bibinfo {volume} {114}},\ \bibinfo {pages}
  {090502} (\bibinfo {year} {2015})}\BibitemShut {NoStop}%
\bibitem [{\citenamefont {Berry}\ \emph {et~al.}(2007)\citenamefont {Berry},
  \citenamefont {Ahokas}, \citenamefont {Cleve},\ and\ \citenamefont
  {Sanders}}]{berry2007efficient}%
  \BibitemOpen
  \bibfield  {author} {\bibinfo {author} {\bibfnamefont {D.~W.}\ \bibnamefont
  {Berry}}, \bibinfo {author} {\bibfnamefont {G.}~\bibnamefont {Ahokas}},
  \bibinfo {author} {\bibfnamefont {R.}~\bibnamefont {Cleve}}, \ and\ \bibinfo
  {author} {\bibfnamefont {B.~C.}\ \bibnamefont {Sanders}},\ }\href {\doibase
  https://doi.org/10.1007/s00220-006-0150-x} {\bibfield  {journal} {\bibinfo
  {journal} {Communications in Mathematical Physics}\ }\textbf {\bibinfo
  {volume} {270}},\ \bibinfo {pages} {359} (\bibinfo {year}
  {2007})}\BibitemShut {NoStop}%
\bibitem [{\citenamefont {Jordan}\ \emph {et~al.}(2012)\citenamefont {Jordan},
  \citenamefont {Lee},\ and\ \citenamefont {Preskill}}]{jordan2012quantum}%
  \BibitemOpen
  \bibfield  {author} {\bibinfo {author} {\bibfnamefont {S.~P.}\ \bibnamefont
  {Jordan}}, \bibinfo {author} {\bibfnamefont {K.~S.}\ \bibnamefont {Lee}}, \
  and\ \bibinfo {author} {\bibfnamefont {J.}~\bibnamefont {Preskill}},\ }\href
  {\doibase 10.1126/science.1217069} {\bibfield  {journal} {\bibinfo  {journal}
  {Science}\ }\textbf {\bibinfo {volume} {336}},\ \bibinfo {pages} {1130}
  (\bibinfo {year} {2012})}\BibitemShut {NoStop}%
\bibitem [{\citenamefont {Low}\ and\ \citenamefont
  {Chuang}(2017)}]{PhysRevLett.118.010501}%
  \BibitemOpen
  \bibfield  {author} {\bibinfo {author} {\bibfnamefont {G.~H.}\ \bibnamefont
  {Low}}\ and\ \bibinfo {author} {\bibfnamefont {I.~L.}\ \bibnamefont
  {Chuang}},\ }\href {\doibase 10.1103/PhysRevLett.118.010501} {\bibfield
  {journal} {\bibinfo  {journal} {Phys. Rev. Lett.}\ }\textbf {\bibinfo
  {volume} {118}},\ \bibinfo {pages} {010501} (\bibinfo {year}
  {2017})}\BibitemShut {NoStop}%
\bibitem [{\citenamefont {Peruzzo}\ \emph {et~al.}(2014)\citenamefont
  {Peruzzo}, \citenamefont {McClean}, \citenamefont {Shadbolt}, \citenamefont
  {Yung}, \citenamefont {Zhou}, \citenamefont {Love}, \citenamefont
  {Aspuru-Guzik},\ and\ \citenamefont {O’brien}}]{peruzzo2014variational}%
  \BibitemOpen
  \bibfield  {author} {\bibinfo {author} {\bibfnamefont {A.}~\bibnamefont
  {Peruzzo}}, \bibinfo {author} {\bibfnamefont {J.}~\bibnamefont {McClean}},
  \bibinfo {author} {\bibfnamefont {P.}~\bibnamefont {Shadbolt}}, \bibinfo
  {author} {\bibfnamefont {M.-H.}\ \bibnamefont {Yung}}, \bibinfo {author}
  {\bibfnamefont {X.-Q.}\ \bibnamefont {Zhou}}, \bibinfo {author}
  {\bibfnamefont {P.~J.}\ \bibnamefont {Love}}, \bibinfo {author}
  {\bibfnamefont {A.}~\bibnamefont {Aspuru-Guzik}}, \ and\ \bibinfo {author}
  {\bibfnamefont {J.~L.}\ \bibnamefont {O’brien}},\ }\href {\doibase
  https://doi.org/10.1038/ncomms5213} {\bibfield  {journal} {\bibinfo
  {journal} {Nature communications}\ }\textbf {\bibinfo {volume} {5}},\
  \bibinfo {pages} {1} (\bibinfo {year} {2014})}\BibitemShut {NoStop}%
\bibitem [{\citenamefont {McClean}\ \emph {et~al.}(2016)\citenamefont
  {McClean}, \citenamefont {Romero}, \citenamefont {Babbush},\ and\
  \citenamefont {Aspuru-Guzik}}]{mcclean2016theory}%
  \BibitemOpen
  \bibfield  {author} {\bibinfo {author} {\bibfnamefont {J.~R.}\ \bibnamefont
  {McClean}}, \bibinfo {author} {\bibfnamefont {J.}~\bibnamefont {Romero}},
  \bibinfo {author} {\bibfnamefont {R.}~\bibnamefont {Babbush}}, \ and\
  \bibinfo {author} {\bibfnamefont {A.}~\bibnamefont {Aspuru-Guzik}},\ }\href
  {\doibase 10.1088/1367-2630/18/2/023023} {\bibfield  {journal} {\bibinfo
  {journal} {New Journal of Physics}\ }\textbf {\bibinfo {volume} {18}},\
  \bibinfo {pages} {023023} (\bibinfo {year} {2016})}\BibitemShut {NoStop}%
\bibitem [{\citenamefont {Lin}\ \emph {et~al.}(2021)\citenamefont {Lin},
  \citenamefont {Dilip}, \citenamefont {Green}, \citenamefont {Smith},\ and\
  \citenamefont {Pollmann}}]{PRXQuantum.2.010342}%
  \BibitemOpen
  \bibfield  {author} {\bibinfo {author} {\bibfnamefont {S.-H.}\ \bibnamefont
  {Lin}}, \bibinfo {author} {\bibfnamefont {R.}~\bibnamefont {Dilip}}, \bibinfo
  {author} {\bibfnamefont {A.~G.}\ \bibnamefont {Green}}, \bibinfo {author}
  {\bibfnamefont {A.}~\bibnamefont {Smith}}, \ and\ \bibinfo {author}
  {\bibfnamefont {F.}~\bibnamefont {Pollmann}},\ }\href {\doibase
  10.1103/PRXQuantum.2.010342} {\bibfield  {journal} {\bibinfo  {journal} {PRX
  Quantum}\ }\textbf {\bibinfo {volume} {2}},\ \bibinfo {pages} {010342}
  (\bibinfo {year} {2021})}\BibitemShut {NoStop}%
\bibitem [{\citenamefont {Yuan}\ \emph {et~al.}(2019)\citenamefont {Yuan},
  \citenamefont {Endo}, \citenamefont {Zhao}, \citenamefont {Li},\ and\
  \citenamefont {Benjamin}}]{yuan2019theory}%
  \BibitemOpen
  \bibfield  {author} {\bibinfo {author} {\bibfnamefont {X.}~\bibnamefont
  {Yuan}}, \bibinfo {author} {\bibfnamefont {S.}~\bibnamefont {Endo}}, \bibinfo
  {author} {\bibfnamefont {Q.}~\bibnamefont {Zhao}}, \bibinfo {author}
  {\bibfnamefont {Y.}~\bibnamefont {Li}}, \ and\ \bibinfo {author}
  {\bibfnamefont {S.~C.}\ \bibnamefont {Benjamin}},\ }\href {\doibase
  10.22331/q-2019-10-07-191} {\bibfield  {journal} {\bibinfo  {journal}
  {{Quantum}}\ }\textbf {\bibinfo {volume} {3}},\ \bibinfo {pages} {191}
  (\bibinfo {year} {2019})}\BibitemShut {NoStop}%
\bibitem [{\citenamefont {Li}\ and\ \citenamefont
  {Benjamin}(2017)}]{li2017efficient}%
  \BibitemOpen
  \bibfield  {author} {\bibinfo {author} {\bibfnamefont {Y.}~\bibnamefont
  {Li}}\ and\ \bibinfo {author} {\bibfnamefont {S.~C.}\ \bibnamefont
  {Benjamin}},\ }\href {\doibase 10.1103/PhysRevX.7.021050} {\bibfield
  {journal} {\bibinfo  {journal} {Phys. Rev. X}\ }\textbf {\bibinfo {volume}
  {7}},\ \bibinfo {pages} {021050} (\bibinfo {year} {2017})}\BibitemShut
  {NoStop}%
\bibitem [{\citenamefont {Endo}\ \emph {et~al.}(2020)\citenamefont {Endo},
  \citenamefont {Sun}, \citenamefont {Li}, \citenamefont {Benjamin},\ and\
  \citenamefont {Yuan}}]{endo2020variational}%
  \BibitemOpen
  \bibfield  {author} {\bibinfo {author} {\bibfnamefont {S.}~\bibnamefont
  {Endo}}, \bibinfo {author} {\bibfnamefont {J.}~\bibnamefont {Sun}}, \bibinfo
  {author} {\bibfnamefont {Y.}~\bibnamefont {Li}}, \bibinfo {author}
  {\bibfnamefont {S.~C.}\ \bibnamefont {Benjamin}}, \ and\ \bibinfo {author}
  {\bibfnamefont {X.}~\bibnamefont {Yuan}},\ }\href {\doibase
  10.1103/PhysRevLett.125.010501} {\bibfield  {journal} {\bibinfo  {journal}
  {Phys. Rev. Lett.}\ }\textbf {\bibinfo {volume} {125}},\ \bibinfo {pages}
  {010501} (\bibinfo {year} {2020})}\BibitemShut {NoStop}%
\bibitem [{\citenamefont {Cirstoiu}\ \emph {et~al.}(2020)\citenamefont
  {Cirstoiu}, \citenamefont {Holmes}, \citenamefont {Iosue}, \citenamefont
  {Cincio}, \citenamefont {Coles},\ and\ \citenamefont
  {Sornborger}}]{cirstoiu2020variational}%
  \BibitemOpen
  \bibfield  {author} {\bibinfo {author} {\bibfnamefont {C.}~\bibnamefont
  {Cirstoiu}}, \bibinfo {author} {\bibfnamefont {Z.}~\bibnamefont {Holmes}},
  \bibinfo {author} {\bibfnamefont {J.}~\bibnamefont {Iosue}}, \bibinfo
  {author} {\bibfnamefont {L.}~\bibnamefont {Cincio}}, \bibinfo {author}
  {\bibfnamefont {P.~J.}\ \bibnamefont {Coles}}, \ and\ \bibinfo {author}
  {\bibfnamefont {A.}~\bibnamefont {Sornborger}},\ }\href {\doibase
  https://doi.org/10.1038/s41534-020-00302-0} {\bibfield  {journal} {\bibinfo
  {journal} {npj Quantum Information}\ }\textbf {\bibinfo {volume} {6}},\
  \bibinfo {pages} {1} (\bibinfo {year} {2020})}\BibitemShut {NoStop}%
\bibitem [{\citenamefont {Benedetti}\ \emph {et~al.}(2021)\citenamefont
  {Benedetti}, \citenamefont {Fiorentini},\ and\ \citenamefont
  {Lubasch}}]{benedetti2021hardware}%
  \BibitemOpen
  \bibfield  {author} {\bibinfo {author} {\bibfnamefont {M.}~\bibnamefont
  {Benedetti}}, \bibinfo {author} {\bibfnamefont {M.}~\bibnamefont
  {Fiorentini}}, \ and\ \bibinfo {author} {\bibfnamefont {M.}~\bibnamefont
  {Lubasch}},\ }\href {\doibase 10.1103/PhysRevResearch.3.033083} {\bibfield
  {journal} {\bibinfo  {journal} {Phys. Rev. Research}\ }\textbf {\bibinfo
  {volume} {3}},\ \bibinfo {pages} {033083} (\bibinfo {year}
  {2021})}\BibitemShut {NoStop}%
\bibitem [{\citenamefont {Barison}\ \emph {et~al.}(2021)\citenamefont
  {Barison}, \citenamefont {Vicentini},\ and\ \citenamefont
  {Carleo}}]{barison2021efficient}%
  \BibitemOpen
  \bibfield  {author} {\bibinfo {author} {\bibfnamefont {S.}~\bibnamefont
  {Barison}}, \bibinfo {author} {\bibfnamefont {F.}~\bibnamefont {Vicentini}},
  \ and\ \bibinfo {author} {\bibfnamefont {G.}~\bibnamefont {Carleo}},\ }\href
  {\doibase 10.22331/q-2021-07-28-512} {\bibfield  {journal} {\bibinfo
  {journal} {{Quantum}}\ }\textbf {\bibinfo {volume} {5}},\ \bibinfo {pages}
  {512} (\bibinfo {year} {2021})}\BibitemShut {NoStop}%
\bibitem [{\citenamefont {Berthusen}\ \emph {et~al.}(2021)\citenamefont
  {Berthusen}, \citenamefont {Trevisan}, \citenamefont {Iadecola},\ and\
  \citenamefont {Orth}}]{berthusen2021quantum}%
  \BibitemOpen
  \bibfield  {author} {\bibinfo {author} {\bibfnamefont {N.~F.}\ \bibnamefont
  {Berthusen}}, \bibinfo {author} {\bibfnamefont {T.~V.}\ \bibnamefont
  {Trevisan}}, \bibinfo {author} {\bibfnamefont {T.}~\bibnamefont {Iadecola}},
  \ and\ \bibinfo {author} {\bibfnamefont {P.~P.}\ \bibnamefont {Orth}},\
  }\href {https://arxiv.org/abs/2112.12654} {\bibfield  {journal} {\bibinfo
  {journal} {arXiv preprint arXiv:2112.12654}\ } (\bibinfo {year}
  {2021})}\BibitemShut {NoStop}%
\bibitem [{\citenamefont {Chan}\ \emph {et~al.}(2023)\citenamefont {Chan},
  \citenamefont {Mu{\~n}oz-Ramo},\ and\ \citenamefont
  {Fitzpatrick}}]{chan2023simulating}%
  \BibitemOpen
  \bibfield  {author} {\bibinfo {author} {\bibfnamefont {H.~H.~S.}\
  \bibnamefont {Chan}}, \bibinfo {author} {\bibfnamefont {D.}~\bibnamefont
  {Mu{\~n}oz-Ramo}}, \ and\ \bibinfo {author} {\bibfnamefont {N.}~\bibnamefont
  {Fitzpatrick}},\ }\href {https://doi.org/10.48550/arXiv.2303.06161}
  {\bibfield  {journal} {\bibinfo  {journal} {arXiv preprint arXiv:2303.06161}\
  } (\bibinfo {year} {2023})}\BibitemShut {NoStop}%
\bibitem [{\citenamefont {Hu}\ \emph {et~al.}(2022)\citenamefont {Hu},
  \citenamefont {Head-Marsden}, \citenamefont {Mazziotti}, \citenamefont
  {Narang},\ and\ \citenamefont {Kais}}]{hu2022general}%
  \BibitemOpen
  \bibfield  {author} {\bibinfo {author} {\bibfnamefont {Z.}~\bibnamefont
  {Hu}}, \bibinfo {author} {\bibfnamefont {K.}~\bibnamefont {Head-Marsden}},
  \bibinfo {author} {\bibfnamefont {D.~A.}\ \bibnamefont {Mazziotti}}, \bibinfo
  {author} {\bibfnamefont {P.}~\bibnamefont {Narang}}, \ and\ \bibinfo {author}
  {\bibfnamefont {S.}~\bibnamefont {Kais}},\ }\href
  {https://doi.org/10.22331/q-2022-05-30-726} {\bibfield  {journal} {\bibinfo
  {journal} {Quantum}\ }\textbf {\bibinfo {volume} {6}},\ \bibinfo {pages}
  {726} (\bibinfo {year} {2022})}\BibitemShut {NoStop}%
\bibitem [{\citenamefont {Han}\ \emph {et~al.}(2021)\citenamefont {Han},
  \citenamefont {Cai}, \citenamefont {Hu}, \citenamefont {Mu}, \citenamefont
  {Ma}, \citenamefont {Xu}, \citenamefont {Wang}, \citenamefont {Wang},
  \citenamefont {Song}, \citenamefont {Zou} \emph
  {et~al.}}]{han2021experimental}%
  \BibitemOpen
  \bibfield  {author} {\bibinfo {author} {\bibfnamefont {J.}~\bibnamefont
  {Han}}, \bibinfo {author} {\bibfnamefont {W.}~\bibnamefont {Cai}}, \bibinfo
  {author} {\bibfnamefont {L.}~\bibnamefont {Hu}}, \bibinfo {author}
  {\bibfnamefont {X.}~\bibnamefont {Mu}}, \bibinfo {author} {\bibfnamefont
  {Y.}~\bibnamefont {Ma}}, \bibinfo {author} {\bibfnamefont {Y.}~\bibnamefont
  {Xu}}, \bibinfo {author} {\bibfnamefont {W.}~\bibnamefont {Wang}}, \bibinfo
  {author} {\bibfnamefont {H.}~\bibnamefont {Wang}}, \bibinfo {author}
  {\bibfnamefont {Y.}~\bibnamefont {Song}}, \bibinfo {author} {\bibfnamefont
  {C.-L.}\ \bibnamefont {Zou}},  \emph {et~al.},\ }\href
  {https://doi.org/10.1103/PhysRevLett.127.020504} {\bibfield  {journal}
  {\bibinfo  {journal} {Physical Review Letters}\ }\textbf {\bibinfo {volume}
  {127}},\ \bibinfo {pages} {020504} (\bibinfo {year} {2021})}\BibitemShut
  {NoStop}%
\bibitem [{\citenamefont {Lindblad}(1976)}]{lindblad1976generators}%
  \BibitemOpen
  \bibfield  {author} {\bibinfo {author} {\bibfnamefont {G.}~\bibnamefont
  {Lindblad}},\ }\href {\doibase https://doi.org/10.1007/BF01608499} {\bibfield
   {journal} {\bibinfo  {journal} {Communications in Mathematical Physics}\
  }\textbf {\bibinfo {volume} {48}},\ \bibinfo {pages} {119} (\bibinfo {year}
  {1976})}\BibitemShut {NoStop}%
\bibitem [{\citenamefont {Gorini}\ \emph {et~al.}(1976)\citenamefont {Gorini},
  \citenamefont {Kossakowski},\ and\ \citenamefont
  {Sudarshan}}]{gorini1976completely}%
  \BibitemOpen
  \bibfield  {author} {\bibinfo {author} {\bibfnamefont {V.}~\bibnamefont
  {Gorini}}, \bibinfo {author} {\bibfnamefont {A.}~\bibnamefont {Kossakowski}},
  \ and\ \bibinfo {author} {\bibfnamefont {E.~C.~G.}\ \bibnamefont
  {Sudarshan}},\ }\href {\doibase https://doi.org/10.1063/1.522979} {\bibfield
  {journal} {\bibinfo  {journal} {Journal of Mathematical Physics}\ }\textbf
  {\bibinfo {volume} {17}},\ \bibinfo {pages} {821} (\bibinfo {year}
  {1976})}\BibitemShut {NoStop}%
\bibitem [{\citenamefont {McLachlan}(1964)}]{mclachlan1964variational}%
  \BibitemOpen
  \bibfield  {author} {\bibinfo {author} {\bibfnamefont {A.}~\bibnamefont
  {McLachlan}},\ }\href {\doibase 10.1080/00268976400100041} {\bibfield
  {journal} {\bibinfo  {journal} {Molecular Physics}\ }\textbf {\bibinfo
  {volume} {8}},\ \bibinfo {pages} {39} (\bibinfo {year} {1964})}\BibitemShut
  {NoStop}%
\bibitem [{\citenamefont {Huang}\ \emph {et~al.}(2020)\citenamefont {Huang},
  \citenamefont {Kueng},\ and\ \citenamefont {Preskill}}]{huang2020predicting}%
  \BibitemOpen
  \bibfield  {author} {\bibinfo {author} {\bibfnamefont {H.-Y.}\ \bibnamefont
  {Huang}}, \bibinfo {author} {\bibfnamefont {R.}~\bibnamefont {Kueng}}, \ and\
  \bibinfo {author} {\bibfnamefont {J.}~\bibnamefont {Preskill}},\ }\href
  {https://doi.org/10.1038/s41567-020-0932-7} {\bibfield  {journal} {\bibinfo
  {journal} {Nature Physics}\ }\textbf {\bibinfo {volume} {16}},\ \bibinfo
  {pages} {1050} (\bibinfo {year} {2020})}\BibitemShut {NoStop}%
\bibitem [{\citenamefont {Nakaji}\ \emph {et~al.}(2022)\citenamefont {Nakaji},
  \citenamefont {Endo}, \citenamefont {Matsuzaki},\ and\ \citenamefont
  {Hakoshima}}]{nakaji2022measurement}%
  \BibitemOpen
  \bibfield  {author} {\bibinfo {author} {\bibfnamefont {K.}~\bibnamefont
  {Nakaji}}, \bibinfo {author} {\bibfnamefont {S.}~\bibnamefont {Endo}},
  \bibinfo {author} {\bibfnamefont {Y.}~\bibnamefont {Matsuzaki}}, \ and\
  \bibinfo {author} {\bibfnamefont {H.}~\bibnamefont {Hakoshima}},\ }\href
  {https://doi.org/10.48550/arXiv.2208.13934} {\bibfield  {journal} {\bibinfo
  {journal} {arXiv preprint arXiv:2208.13934}\ } (\bibinfo {year}
  {2022})}\BibitemShut {NoStop}%
\bibitem [{\citenamefont {Clinton}\ \emph {et~al.}(2021)\citenamefont
  {Clinton}, \citenamefont {Bausch},\ and\ \citenamefont
  {Cubitt}}]{clinton2021hamiltonian}%
  \BibitemOpen
  \bibfield  {author} {\bibinfo {author} {\bibfnamefont {L.}~\bibnamefont
  {Clinton}}, \bibinfo {author} {\bibfnamefont {J.}~\bibnamefont {Bausch}}, \
  and\ \bibinfo {author} {\bibfnamefont {T.}~\bibnamefont {Cubitt}},\ }\href
  {https://doi.org/10.1038/s41467-021-25196-0} {\bibfield  {journal} {\bibinfo
  {journal} {Nature communications}\ }\textbf {\bibinfo {volume} {12}},\
  \bibinfo {pages} {1} (\bibinfo {year} {2021})}\BibitemShut {NoStop}%
\bibitem [{\citenamefont {Vatan}\ and\ \citenamefont
  {Williams}(2004)}]{PhysRevA.69.032315}%
  \BibitemOpen
  \bibfield  {author} {\bibinfo {author} {\bibfnamefont {F.}~\bibnamefont
  {Vatan}}\ and\ \bibinfo {author} {\bibfnamefont {C.}~\bibnamefont
  {Williams}},\ }\href {\doibase 10.1103/PhysRevA.69.032315} {\bibfield
  {journal} {\bibinfo  {journal} {Phys. Rev. A}\ }\textbf {\bibinfo {volume}
  {69}},\ \bibinfo {pages} {032315} (\bibinfo {year} {2004})}\BibitemShut
  {NoStop}%
\bibitem [{\citenamefont {Vidal}\ and\ \citenamefont
  {Dawson}(2004)}]{PhysRevA.69.010301}%
  \BibitemOpen
  \bibfield  {author} {\bibinfo {author} {\bibfnamefont {G.}~\bibnamefont
  {Vidal}}\ and\ \bibinfo {author} {\bibfnamefont {C.~M.}\ \bibnamefont
  {Dawson}},\ }\href {\doibase 10.1103/PhysRevA.69.010301} {\bibfield
  {journal} {\bibinfo  {journal} {Phys. Rev. A}\ }\textbf {\bibinfo {volume}
  {69}},\ \bibinfo {pages} {010301} (\bibinfo {year} {2004})}\BibitemShut
  {NoStop}%
\bibitem [{\citenamefont {Kraus}\ and\ \citenamefont
  {Cirac}(2001)}]{PhysRevA.63.062309}%
  \BibitemOpen
  \bibfield  {author} {\bibinfo {author} {\bibfnamefont {B.}~\bibnamefont
  {Kraus}}\ and\ \bibinfo {author} {\bibfnamefont {J.~I.}\ \bibnamefont
  {Cirac}},\ }\href {\doibase 10.1103/PhysRevA.63.062309} {\bibfield  {journal}
  {\bibinfo  {journal} {Phys. Rev. A}\ }\textbf {\bibinfo {volume} {63}},\
  \bibinfo {pages} {062309} (\bibinfo {year} {2001})}\BibitemShut {NoStop}%
\bibitem [{\citenamefont {Daskin}\ and\ \citenamefont
  {Kais}(2017)}]{daskin2017ancilla}%
  \BibitemOpen
  \bibfield  {author} {\bibinfo {author} {\bibfnamefont {A.}~\bibnamefont
  {Daskin}}\ and\ \bibinfo {author} {\bibfnamefont {S.}~\bibnamefont {Kais}},\
  }\href {\doibase https://doi.org/10.1007/s11128-016-1452-3} {\bibfield
  {journal} {\bibinfo  {journal} {Quantum Information Processing}\ }\textbf
  {\bibinfo {volume} {16}},\ \bibinfo {pages} {1} (\bibinfo {year}
  {2017})}\BibitemShut {NoStop}%
\bibitem [{\citenamefont {Terashima}\ and\ \citenamefont
  {Ueda}(2005)}]{terashima2005nonunitary}%
  \BibitemOpen
  \bibfield  {author} {\bibinfo {author} {\bibfnamefont {H.}~\bibnamefont
  {Terashima}}\ and\ \bibinfo {author} {\bibfnamefont {M.}~\bibnamefont
  {Ueda}},\ }\href {\doibase https://doi.org/10.1142/S0219749905001456}
  {\bibfield  {journal} {\bibinfo  {journal} {International Journal of Quantum
  Information}\ }\textbf {\bibinfo {volume} {3}},\ \bibinfo {pages} {633}
  (\bibinfo {year} {2005})}\BibitemShut {NoStop}%
\bibitem [{\citenamefont {Berthusen}\ \emph {et~al.}(2022)\citenamefont
  {Berthusen}, \citenamefont {Trevisan}, \citenamefont {Iadecola},\ and\
  \citenamefont {Orth}}]{PhysRevResearch.4.023097}%
  \BibitemOpen
  \bibfield  {author} {\bibinfo {author} {\bibfnamefont {N.~F.}\ \bibnamefont
  {Berthusen}}, \bibinfo {author} {\bibfnamefont {T.~V.}\ \bibnamefont
  {Trevisan}}, \bibinfo {author} {\bibfnamefont {T.}~\bibnamefont {Iadecola}},
  \ and\ \bibinfo {author} {\bibfnamefont {P.~P.}\ \bibnamefont {Orth}},\
  }\href {\doibase 10.1103/PhysRevResearch.4.023097} {\bibfield  {journal}
  {\bibinfo  {journal} {Phys. Rev. Research}\ }\textbf {\bibinfo {volume}
  {4}},\ \bibinfo {pages} {023097} (\bibinfo {year} {2022})}\BibitemShut
  {NoStop}%
\bibitem [{\citenamefont {Ekert}\ \emph {et~al.}(2002)\citenamefont {Ekert},
  \citenamefont {Alves}, \citenamefont {Oi}, \citenamefont {Horodecki},
  \citenamefont {Horodecki},\ and\ \citenamefont {Kwek}}]{ekert2002direct}%
  \BibitemOpen
  \bibfield  {author} {\bibinfo {author} {\bibfnamefont {A.~K.}\ \bibnamefont
  {Ekert}}, \bibinfo {author} {\bibfnamefont {C.~M.}\ \bibnamefont {Alves}},
  \bibinfo {author} {\bibfnamefont {D.~K.~L.}\ \bibnamefont {Oi}}, \bibinfo
  {author} {\bibfnamefont {M.}~\bibnamefont {Horodecki}}, \bibinfo {author}
  {\bibfnamefont {P.}~\bibnamefont {Horodecki}}, \ and\ \bibinfo {author}
  {\bibfnamefont {L.~C.}\ \bibnamefont {Kwek}},\ }\href {\doibase
  10.1103/PhysRevLett.88.217901} {\bibfield  {journal} {\bibinfo  {journal}
  {Phys. Rev. Lett.}\ }\textbf {\bibinfo {volume} {88}},\ \bibinfo {pages}
  {217901} (\bibinfo {year} {2002})}\BibitemShut {NoStop}%
\bibitem [{\citenamefont {Buhrman}\ \emph {et~al.}(2001)\citenamefont
  {Buhrman}, \citenamefont {Cleve}, \citenamefont {Watrous},\ and\
  \citenamefont {de~Wolf}}]{buhrman2001quantum}%
  \BibitemOpen
  \bibfield  {author} {\bibinfo {author} {\bibfnamefont {H.}~\bibnamefont
  {Buhrman}}, \bibinfo {author} {\bibfnamefont {R.}~\bibnamefont {Cleve}},
  \bibinfo {author} {\bibfnamefont {J.}~\bibnamefont {Watrous}}, \ and\
  \bibinfo {author} {\bibfnamefont {R.}~\bibnamefont {de~Wolf}},\ }\href
  {\doibase 10.1103/PhysRevLett.87.167902} {\bibfield  {journal} {\bibinfo
  {journal} {Phys. Rev. Lett.}\ }\textbf {\bibinfo {volume} {87}},\ \bibinfo
  {pages} {167902} (\bibinfo {year} {2001})}\BibitemShut {NoStop}%
\bibitem [{\citenamefont {Foulds}\ \emph {et~al.}(2021)\citenamefont {Foulds},
  \citenamefont {Kendon},\ and\ \citenamefont
  {Spiller}}]{foulds2021controlled}%
  \BibitemOpen
  \bibfield  {author} {\bibinfo {author} {\bibfnamefont {S.}~\bibnamefont
  {Foulds}}, \bibinfo {author} {\bibfnamefont {V.}~\bibnamefont {Kendon}}, \
  and\ \bibinfo {author} {\bibfnamefont {T.}~\bibnamefont {Spiller}},\ }\href
  {\doibase 10.1088/2058-9565/abe458} {\bibfield  {journal} {\bibinfo
  {journal} {Quantum Science and Technology}\ }\textbf {\bibinfo {volume}
  {6}},\ \bibinfo {pages} {035002} (\bibinfo {year} {2021})}\BibitemShut
  {NoStop}%
\bibitem [{\citenamefont {Lawson}\ and\ \citenamefont
  {Hanson}(1974)}]{lawson1974solving}%
  \BibitemOpen
  \bibfield  {author} {\bibinfo {author} {\bibfnamefont {C.~L.}\ \bibnamefont
  {Lawson}}\ and\ \bibinfo {author} {\bibfnamefont {R.~J.}\ \bibnamefont
  {Hanson}},\ }\href@noop {} {\  (\bibinfo {year} {1974})}\BibitemShut
  {NoStop}%
\bibitem [{\citenamefont {Hansen}\ and\ \citenamefont
  {O’Leary}(1993)}]{hansen1993use}%
  \BibitemOpen
  \bibfield  {author} {\bibinfo {author} {\bibfnamefont {P.~C.}\ \bibnamefont
  {Hansen}}\ and\ \bibinfo {author} {\bibfnamefont {D.~P.}\ \bibnamefont
  {O’Leary}},\ }\href {\doibase https://doi.org/10.1137/0914086} {\bibfield
  {journal} {\bibinfo  {journal} {SIAM journal on scientific computing}\
  }\textbf {\bibinfo {volume} {14}},\ \bibinfo {pages} {1487} (\bibinfo {year}
  {1993})}\BibitemShut {NoStop}%
\bibitem [{\citenamefont {Hansen}(1992)}]{hansen1992analysis}%
  \BibitemOpen
  \bibfield  {author} {\bibinfo {author} {\bibfnamefont {P.~C.}\ \bibnamefont
  {Hansen}},\ }\href {\doibase https://doi.org/10.1137/1034115} {\bibfield
  {journal} {\bibinfo  {journal} {SIAM review}\ }\textbf {\bibinfo {volume}
  {34}},\ \bibinfo {pages} {561} (\bibinfo {year} {1992})}\BibitemShut
  {NoStop}%
\end{thebibliography}%
\end{document}